\documentclass[letterpaper,11pt]{article}
\usepackage{fullpage}
\usepackage{times}
\usepackage{soul}
\usepackage{url}
\usepackage[hidelinks,colorlinks,allcolors=magenta]{hyperref}
\usepackage[utf8]{inputenc}
\usepackage{lmodern}
\usepackage[small]{caption}
\usepackage{graphicx}
\usepackage{amsmath}
\usepackage{amsthm}
\usepackage{booktabs}
\usepackage{algorithm}
\usepackage{algorithmic}
\usepackage[switch]{lineno}
\usepackage[draft]{changes}
\usepackage[numbers,square,sort]{natbib}
\newtheorem{theorem}{Theorem}

\newtheorem{lemma}{Lemma}
\usepackage{amsmath,amssymb,amsthm}
\usepackage{mathtools}
\usepackage{xfrac}
\usepackage{graphicx}
\usepackage{todonotes}
\usepackage{xcolor}
\usepackage{cleveref}
\newtheorem{corollary}{Corollary}

\theoremstyle{definition}
\newtheorem{definition}{Definition}
\usepackage{float}
\usepackage{graphicx}
\usepackage{subcaption}

\newcommand*\dif{\mathop{}\!\mathrm{d}}
\newcommand{\bmid}{\ \Big|\ }

\DeclarePairedDelimiter{\floor}{\lfloor}{\rfloor}
\DeclarePairedDelimiter{\brac}{\Big[}{\Big]}
\renewcommand{\ge}{\geqslant}
\renewcommand{\geq}{\geqslant}
\renewcommand{\le}{\leqslant}
\renewcommand{\leq}{\leqslant}
\newcommand{\set}[1]{\left\{#1\right\}}
\newcommand{\E}{\mathbb{E}}
\newcommand{\bbN}{\mathbb{N}}
\newcommand{\bbR}{\mathbb{R}}
\newcommand{\calQ}{\mathcal{Q}}
\newcommand{\calS}{\mathcal{S}}
\newcommand{\calD}{\mathcal{D}}
\newcommand{\bucket}{B}

\newcommand{\same}{\operatorname{same}}
\newcommand{\opp}{\operatorname{opp}}
\newcommand{\low}{\textrm{low}}
\newcommand{\high}{\textrm{high}}
\newcommand{\inc}{\textrm{inc}}
\newcommand{\dec}{\textrm{dec}}

\newcommand{\ulg}{\operatorname{ULG}}
\newcommand{\ns}{\operatorname{NS}}
\newcommand{\perf}{\operatorname{perf}}

\usepackage{pgfplots}
    \pgfplotsset{
        compat=1.3,
    }
    \pgfplotscreateplotcyclelist{line styles}{
        black,solid\\
        blue,dashed\\
        red,dotted\\
        orange,dashdotted\\
    }

\title{What is Best for Students, Numerical Scores or Letter Grades?}

\author{
Evi Micha$^1$\and
Shreyas Sekar$^{2,3}$\and
Nisarg Shah$^1$
}
\date{$^1$Department of Computer Science, University of Toronto\\
$^2$Rotman School of Management, University of Toronto\\
$^3$Department of Management, University of Toronto Scarborough\\
\{emicha, nisarg\}@cs.toronto.edu,
shreyas.sekar@rotman.utoronto.ca
}
\begin{document}\allowdisplaybreaks

\maketitle
\begin{abstract}
We study letter grading schemes, which are routinely employed for evaluating student performance. Typically, a numerical score obtained via one or more evaluations is converted into a letter grade (e.g., A+, B-, etc.) by associating a disjoint interval of numerical scores to each letter grade. 

We propose the first model for studying the (de)motivational effects of such grading on the students and, consequently, on their performance in future evaluations. We use the model to compare uniform letter grading schemes, in which the range of scores is divided into equal-length parts that are mapped to the letter grades, to numerical scoring, in which the score is not converted to any letter grade (equivalently, every score is its own letter grade).  

Theoretically, we identify realistic conditions under which numerical scoring is better than any uniform letter grading scheme. Our experiments confirm that this holds under even weaker conditions, but also find cases where the converse occurs.
\end{abstract}

\section{Introduction}\label{sec:intro}

Student evaluations and grading play an integral and influential role in every individual's academic experience. Naturally, there has been widespread debate among researchers and policy-makers about the efficacy of various grading systems such as \emph{letter v.s. number grades}. For instance, coarse-grained grading schemes (i.e., letter grades) are considered to be less noisy indicators of performance, and stronger signals of status, and consequently, are the norm in North American universities. At the same time, there is also growing awareness that the grade itself affects performance independent of student ability, i.e., the grades are ``not just an output of the educational process, they may also be an input"~\cite{gray2022effect}. For example, empirical evidence suggests the disclosure of midterm grades may motivate or demotivate students to perform better in a future exam, controlling for other effects. In light of this evidence, it is clear that the design of a grading system must be a deliberate choice that takes into account student welfare in addition to other extraneous factors~\cite{guskey2011five}.
 In this work, we take an analytical approach and study the design of an optimal grading system with a particular focus on numeric v.s. uniform letter grades.\footnote{We use the term uniform letter grades to refer to letter grading schemes where each letter grade corresponds to an equal sized score range, e.g., $[90,100]\rightarrow$A+, $[80,90]\rightarrow$A-, and so on. } As far as we are aware, this work is among the first to look at the problem of designing a grading scheme with the explicit objective of improving student performance in future tests. Our model captures the impact of grades on future performance via two well-motivated effects:

 \begin{enumerate}
   \item \textbf{Anchoring:} In any given test, students anchor themselves to (i.e., in expectation perform as well as) a specific score or performance level based on their intrinsic ability. We refer to this anchor as the intrinsic quality.
    
    \item \textbf{(De)Motivation}: When the students' actual score falls above (below) their intrinsic quality, they get (de)motivated and subsequently, their expectation increases (decreases) for future tests. This is a phenomenon that has been widely noticed in practice~\cite{deci1999meta,dev1997intrinsic,cameron1994reinforcement}.
 \end{enumerate}

In this regard, our work departs from other papers in this area, where students are often modelled as status-maximizers~\cite{dubey2010grading}, i.e., their intrinsic motivation for a better grade stems from a desire to rank above their fellow students. Our model does not induce any artificial scarcity (status) and instead the fundamental friction is a result of noisy performance and how the same grading rule affects different students differently.

To better illustrate how different grading schemes impact student performance under our model, consider a student with an intrinsic quality of $q_1 = 85$. Suppose that the student scores $s_1 = 81$ in the midterm exam. Disclosing this numeric score may demotivate the student, which may reduce her effective intrinsic quality for the final exam. This adverse effect may be prevented if a (coarser) letter grading scheme is used, in which (say) all students (including the student under consideration) whose scores lie in $[80,90]$ are assigned a letter grade of $A-$. However, consider another student whose intrinsic quality is $q_2=91$ and whose midterm score is $s_2=89$. Receiving the same letter grade $A-$ as everyone who scored in $[80,90]$ may be more demotivating to her than receiving her numerical score of $s_2 = 89$. Hence, the overall effect of using a letter grading scheme remains unclear.

There is another subtle issue to be considered. While the comparison made by a student between her numerical score and intrinsic true quality is straightforward,\footnote{This assumes that students is aware of their own intrinsic true qualities, but it may suffice for them to have noisy estimates.} it is not obvious how a student should compare her intrinsic true quality to a letter grade received (such as $A-$). This depends on how the student perceives the letter grade. To that end, we use a scheme for mapping letter grades back to representative numerical scores: each letter grade is mapped to the \emph{midpoint} of the interval containing all the scores that were mapped to that letter grade. For example, if all scores in the range $[80,90]$ are mapped to the letter grade $A-$, then $A-$ is mapped back to (i.e., considered worth) a score of $85$, which is what any student receiving $A-$ would compare to her intrinsic quality. 

This \emph{midpoint scheme} has three in-built advantages. First, it reflects how the letter grades may truly be perceived in the outside world (and thus by the students) as it is actually used in the real world~\cite{midpoint1,midpoint2}. Second, the association of a letter grade to the midpoint of its score range accurately conveys the (average) performance of a student receiving that letter grade. Third, in the absence of any structure on how the grades are perceived, the question we ask in this work --- which letter grading schemes would lead to the maximum average student performance? --- would have a trivial and rather unsatisfactory answer: assign all students a grade worth $100$ to maximally motivate them. The midpoint scheme makes this impossible: if the same grade (say $A+$) was assigned to all the students with scores anywhere in $[0,100]$, it would only be worth $50$.

Building on the ideas presented in this example, we develop a framework to compare various grading systems in an environment with \emph{sequential testing}. This includes evaluations within a course, e.g., a midterm followed by a final exam, but also grading across related courses, e.g., a student taking Calculus 101 followed by Calculus 102. Since a student's intrinsic quality increases after a test if 
the grade received is higher than her intrinsic quality and decreases otherwise,  our aim is 
\begin{quote}
{\em to compare different grading schemes and choose the one that provides a higher quality improvement (or a lower quality degradation).   }
\end{quote}

\smallskip\noindent\textbf{Our results.} In this work, we compare the numerical scoring scheme, where the student learns her exact score in an evaluation, to uniform letter grading schemes, where the interval of scores is partitioned into $T$ equal-length intervals mapping to different letter grades (and each interval is represented by its midpoint). While uniform letter grading is not completely realistic,
we view our work as a starting point for the curiously unaddressed problem of quantitatively optimizing letter grading schemes and a stepping stone for future work to build on. That said, we note that real-world letter grading schemes (at least those used in North American universities) are close to uniform, once a very large interval mapped to the failing grade and a somewhat large interval mapped to very top grade are omitted. Since very few students fall in these two intervals, this omission does not significantly affect the overall analysis.

First, we theoretically study the case where two sequential evaluations take place, such as midterm and final exam. We show that under natural conditions, numerical scoring and all uniform letter grading schemes have equal performance when the motivational and demotivational effects are equally strong, and otherwise, either numerical scoring outperforms all uniform letter grading schemes or the opposite happens. Analytically identifying when each scheme outperforms the other turns out to be far from obvious and subtly dependent on properties of the distributions of intrinsic true qualities and scores, even for this limited setting. Using carefully constructed bijections between students, we are able to identify additional conditions under which numerical scoring outperforms all uniform letter grading schemes when the demotivational effect is stronger than the motivational effect, and the opposite happens when the demotivational effect is weaker than the motivational effect. Since there is significant evidence that negative events have a greater impact than positive events~\cite{baumeister2001bad,coleman1987students}, we expect the demotivational effect to be stronger than the motivational effect; thus, our results are in favour of numerical scoring.

Next, we empirically compare numerical scoring to uniform letter grading schemes. Under two sequential evaluations, we observe that numerical scoring continues to outperform uniform letter grading when the demotivational effect is stronger (and the opposite continues to hold when the motivational effect is stronger), even under more realistic conditions than in our theoretical analysis, such as when the true qualities of the students follow a (truncated) normal distribution. However, surprisingly, when more than two evaluations take place, the effect is reversed. Even after just six sequential evaluations, uniform letter grading begins to outperform numerical scoring when the demotivational effect is stronger (and the opposite holds when the motivational effect is stronger). In the intermediate stage between these two regimes, there is another surprising effect: with four sequential evaluations, numerical scoring outperforms uniform letter grading regardless of which effect is stronger! 

Our results indicate that the choice of the grading scheme depends on the application at hand: with fewer evaluations (e.g., courses with just a few tests or shorter education programs with just a few semesters), numerical scoring may be better, while with many evaluations (e.g., courses with weekly tests or longer education programs), uniform letter grading may be better.  At a high level, although our work draws on literature from fields such as economics and psychology, it provides a fundamental perspective on the question of student grading within the framework of multi-agent systems, i.e., where each student is modeled as an agent whose behavior depends on the decisions made by the system. Our results open up the possibility of designing grading systems that are easy to implement, approximately-optimal, and take into account students' incentives.

\smallskip\noindent\textbf{Related work.} There is a rich literature on comparing grading schemes using various objectives. However, to the best of our knowledge, none of these papers study the objective of improving student quality that we focus on. 

Several works have studied, both theoretically and empirically, how the effort exerted by students for an evaluation depends on the grading scheme to be used~\cite{parades17,BROWNBACK2018113,Main14,CZIBOR2020}. For example, when using pass/fail grading, a student may try hard enough to pass (with high probability), but not any harder. Our work is orthogonal to this: we focus on effect of the outcome of one evaluation on the student motivation in \emph{subsequent} evaluations. 

Another related work is that of \citet{Sikora2015}, who also compares grading schemes, but his goal is to study the tradeoff between conveying the most information about the student's true quality and minimizing noise due to factors unrelated to the true quality, not the (de)motivational effects of the grading scheme in subsequent evaluations. In our work, the task of keeping the grades ``consistent'' with the actual performance is indirectly performed by the midpoint scheme. 

\citet{db2006} and \citet{BSJ2020} also study how the grading scheme used may impact students' psychological well-being and stress levels, but do not focus on the impact of this in subsequent evaluations. 

\section{Model}\label{sec:model}

Define $[k] = \set{1,\ldots,k}$ for $k \in \bbN$. We introduce a model in which the grading scheme used in one evaluation can motivate or demotivate students, affecting their performance in future evaluations.  

\smallskip\noindent\textbf{True qualities.} A student begins with an intrinsic (true) quality $q$ drawn from a (nonatomic) prior $\calQ$ with probability density function (PDF) $f_{\calQ}(\cdot)$. For simplicity, let the support of $\calQ$ be $[0,1]$.

\smallskip\noindent\textbf{Scores.} There is a \emph{score model} $\calS$ such that the numerical performance (score) of a student with true quality $q$ in the first evaluation, denoted $s\in [0,1]$, is drawn from the (nonatomic) distribution $\calS(q)$ with PDF $f_{\calS}(\cdot;q)$. We focus on score models in which the expected score of a student is equal to their true quality, i.e., $\E_{s \sim \calS(q)} [s] = q$ for all $q \in [0,1]$. 

\smallskip\noindent\textbf{Grades.} A grading scheme is a function $\bucket: [0,1] \to [0,1]$ that maps the score to a grade. 

\emph{Letter grading.} A \emph{letter grading} scheme $\bucket_{\vec{c}}$ is specified by a vector $\vec{c} = (c_0=0,c_1,\ldots,c_{T-1},c_T=1)$, for some $T \in \bbN$ (referred to as the number of grades) and $c_i \ge c_{i-1}$ for all $i \in [T]$, and is given by $\bucket_{\vec{c}}(s) = \frac{c_{i-1}+c_i}{2}$ for all $i \in [T]$ and $s \in [c_{i-1},c_i)$. That is, it partitions $[0,1)$ into finitely many disjoint intervals (one for each grade) and maps a score to the midpoint of the interval containing it. 

\emph{Uniform letter grading.} We are particularly interested in the \emph{uniform letter grading} ($\ulg$) scheme. For a given number of grades $T \in \bbN$, uniform letter grading with $T$ grades, denoted $\ulg_T$, is specified by $c_i = i/T$ for each $i \in [T]$. In other words, it partitions $[0,1)$ into $T$ equal-length intervals. We will use $\Delta(T) = 1/T$ to denote the length of the interval, dropping $T$ from the argument when it is clear from the context. Formally, we have that for all $s \in [0,1)$,\footnote{Because we assume nonatomic distributions, it does not matter what $\ulg_T(1)$ is. We will use the convention that $\ulg_T(1)=1$.}
\[
	\ulg_T(s) = \left(\floor{\sfrac{s}{\Delta}}+\sfrac{1}{2}\right) \cdot \Delta.
\]
For instance, $\ulg_{10}$ maps all scores in $[0,0.1)$ to $0.05$, all scores in $[0.1,0.2)$ to $0.15$, and so on. We restrict our focus to uniform grading schemes for two reasons: a) it is straightforward and easy to implement in practice; b) given that different institutions following different grading schemes, this allows us to broadly compare letter and number grading without getting lost in the minutiae. Further our assumption that each letter grade maps to the midpoint of an interval is common practice across universities~\cite{midpoint1,midpoint2} as well as the literature~\cite{mcewan2021grade,nisbet1975adding}. More generally, it is consistent with the practice of assigning a score or grade-point to each letter grade.

\emph{Numerical scoring.} We will compare (uniform) letter grading to \emph{numerical scoring} ($\ns$), given by $\ns(s) = s$ for all $s \in [0,1]$. Under numerical scoring, scores are not rounded to any grades. This can also be viewed as the limit of uniform letter grading with $T \to \infty$ grades. 

\smallskip\noindent\textbf{(De)motivation.} The grades affect students' level of motivation in subsequent evaluations. Under grading scheme $\bucket$, a student compares their true quality $q$ to the obtained grade $\bucket(s)$. If the grade is higher than the true quality, the student experiences a motivational boost, but in the converse case, gets demotivated. We model this by assuming that the effective true quality of the student for the next evaluation changes to $q' = q + h(q,\bucket(s))$, where
\[
        h(q,\bucket(s)) =
        \begin{cases}
        \alpha_m \cdot (\bucket(s)-q), &\text{if $\bucket(s) \ge q$},\\
        - \alpha_d \cdot (q-\bucket(s)), &\text{if $\bucket(s) < q$}.
        \end{cases}
\]
We refer to $\alpha_m,\alpha_d \in \bbR_{\ge 0}$ as \emph{motivation and demotivation coefficients}, respectively. Note that the amount of (de)motivation is proportional to the difference between the obtained grade and the true quality. In the next evaluation, the student  obtains a score $s'$ drawn from $\calS(q')$. We remark that when $\alpha_m,\alpha_d \in [0,1]$, we automatically have $q' \in [0,1]$; thus, we focus on this range of parameters.\footnote{In principle, one can also use larger coefficients and truncate $q'$ to lie in $[0,1]$.} Our choice of a linear model for demotivation follows from studies showing that student performance is linearly dependent on both external~\cite{christensen1998linear} and internal stimuli~\cite{latham2007new}. Additionally, even when the actual behaviour is more complex, our model serves as a first-order approximation when $(B(s)-q)$ is small.

\smallskip\noindent\textbf{Goal.}
Intuitively, we are interested in choosing grading schemes that achieve a higher increase (or a lower decrease) in the average student quality. Thus, we define the {\em performance} of a grading scheme $B$ as:
\begin{align*}
    \perf(\bucket) \triangleq \E_{q \sim \calQ, s \sim \calS(q)}[q'-q] 
\end{align*}
where $q'=q+h(q,\bucket(s))$. Due to linearity of expectation, 
\[
\perf(\bucket) = \E_{q \sim \calQ, s \sim \calS(q)}[q'-q]= \E_{q \sim \calQ, s \sim \calS(q)}[h(q,\bucket(s))].
\]
Thus, we compare $\E_{q \sim \calQ, s \sim \calS(q)}[h(q,\bucket(s))]$ under numerical scoring and uniform letter grading. Hereinafter, we omit $q \sim \calQ$ and $s \sim \calS(q)$ from an expression of expectation, whenever it is clear from the context. 

Note that for our theoretical analysis, we focus on the case of two evaluations. Later, we empirically study the case of more than two evaluations. 

\section{Theoretical Results}\label{sec:theory}

In this section, we derive theoretical results for the performance of uniform letter grading schemes and numerical scoring, when students participate in two sequential evaluations and identify conditions under which numerical scoring outperforms every uniform letter grading scheme, and conditions under which the converse holds. Let us begin by introducing two useful definitions.

\begin{definition}[Jointly Symmetric Distributions]
We say that the true quality prior $\calQ$ and the score model $\calS$ are \emph{jointly symmetric} if $f_{\calQ}(q) \cdot f_{\calS}(s;q) = f_{\calQ}(1-q) \cdot f_{\calS}(1-s;1-q)$ for all $s,q \in [0,1]$. 
\end{definition}

Joint symmetry requires that true qualities and scores are symmetric across $[0,1]$. That is, the probability of having true quality $q$ and receiving score $s$ should be the same as the probability of having true quality $1-q$ and receiving score $1-s$. If the true quality prior is uniform, then this means the score distribution $\calS(q)$ should be the mirror image of the score distribution $\calS(1-q)$. Note that joint symmetry does not necessarily require symmetry of the ``noise'' contained in the score compared to the true quality. For example, we do not need $f_{\calS}(s=0.4;q=0.5) = f_{\calS}(s=0.6;q=0.5)$. 

\begin{definition}[Symmetric Grading Scheme]
We say that a grading scheme $\bucket$ is \emph{symmetric} if $\bucket(1-s) = 1-\bucket(s)$ for all $s \in [0,1]$.
\end{definition}

The reader can check that numerical scoring ($\ns$) and uniform letter grading schemes ($\ulg_T$ for any $T \in \bbN$) are symmetric. Our first result shows that under such symmetry, the performance of the grading scheme is linear in the difference between the motivation and demotivation coefficients. As we later show in \Cref{cor:weak-symm}, this allows us to compare numerical scoring to uniform letter grading. 

\begin{theorem}\label{thm:weak-symm}
When the true quality prior $\calQ$ and the score model $\calS$ are jointly symmetric, and the grading scheme $\bucket$ is symmetric, then we have
\begin{equation}\label{eqn:weak-symm-perf}
	\perf(\bucket) = \frac{\alpha_m-\alpha_d}{2} \cdot \E_{q \sim \calQ, s \sim \calS(q)} \brac{|q-\bucket(s)|}.
\end{equation}
\end{theorem}
\begin{proof}

Note that due to $\calQ$ and $\calS$ being jointly symmetric, the pairs $(q,s)$ and $(1-q,1-s)$ are sampled with equal density. Hence, we have that
\begin{equation}\label{eqn:main}
    \E\brac{h(q,\bucket(s))} =\frac{1}{2} \cdot \E\brac{h(q,\bucket(s))+h(1-q,\bucket(1-s))}.
\end{equation}

Due to the symmetry of the grading scheme, we have $\bucket(1-s) = 1-\bucket(s)$, which implies that the two terms $h(q,\bucket(s))$ and $h(1-q,\bucket(1-s))$ are motivation and demotivation by the same amount. Hence, 
{\small\begin{align*}
    \E\brac{h(q,\bucket(s))+h(1-q,\bucket(1-s))} = (\alpha_m-\alpha_d) \cdot \E
    \brac{|q-\bucket(s)|}. 
\end{align*}}

Plugging this into \Cref{eqn:main}, we get the result. 
\end{proof}

\begin{corollary}\label{cor:weak-symm}
Assume that the true quality prior $\calQ$ and the score model $\calS$ are jointly symmetric. Then, all symmetric grading schemes have equal performance if $\alpha_m = \alpha_d$. Further, if $\alpha_m \neq \alpha_d$, for every $T \in \bbN$ one of the following conditions holds.
\begin{enumerate}
    \item\label{cor:real} Uniform letter grading with $T$ grades is at least as good as numerical scoring if $\alpha_m > \alpha_d$, and the converse holds if $\alpha_m < \alpha_d$.
    \item\label{cor:fake} Uniform letter grading with $T$ grades is at least as good as numerical scoring if $\alpha_m < \alpha_d$, and the converse holds if $\alpha_m > \alpha_d$.
\end{enumerate}
\end{corollary}
\begin{proof}
The first claim regarding $\alpha_m = \alpha_d$ follows immediately from \Cref{eqn:weak-symm-perf}. For the second claim regarding $\alpha_m \neq \alpha_d$, note that the comparison between numerical scoring and uniform letter grading with $T$ buckets reduces to the sign of $\E[|q-\ns(s)|-|q-\ulg_T(s)|]$, and depending on this sign, one of the two statements in the corollary holds. 
\end{proof}

\Cref{cor:weak-symm} tells us that having equal motivation and demotivation coefficients ($\alpha_m = \alpha_d$) is the turning point: between uniform letter grading with a fixed number of grades and numerical scoring, one is better when $\alpha_m < \alpha_d$ but the other becomes better when $\alpha_m > \alpha_d$. But it does not tell us \emph{which} one is better in each case. 

Our next result identifies a sufficient condition under which this dilemma is settled: uniform letter grading is better when $\alpha_m > \alpha_d$ and numerical scoring is better when $\alpha_m < \alpha_d$. To introduce this sufficient condition, we need to define the following natural property of the score model. 

\begin{definition}[Ex-Ante Single-Peaked Score Model]
We say that the score model $\calS$ is \emph{ex-ante single-peaked} if, for every $q \in [0,1]$, $f_{\calS}(\cdot;q)$ is single-peaked with the peak at $q$, i.e., $f_{\calS}(s;q) \le f_{\calS}(s';q)$ for all $s \le s' \le q$ and $s \ge s' \ge q$.
\end{definition}

Intuitively, in an ex-ante single-peaked score model, scores closer to the true quality are more likely than scores farther from the true quality. 

For a fixed $T$, we also denote with $\calD$ the set of all pairs of true qualities and scores that belong to the same letter grade interval, i.e., $\calD = \set{(q,s) : \ulg_T(q)=\ulg_T(s)}$. For example, if $T=10$, $(q=0.51,s=0.59) \in \calD$ but $(q=0.51,s'=0.49) \notin \calD$. 

\begin{theorem}\label{thm:weak-symm-impr}
Fix any $T \in \bbN$.  Assume that the true quality prior $\calQ$ and the score model $\calS$ satisfy the following.
\begin{enumerate}
    \item $\calQ$ and $\calS$ are jointly symmetric;
    \item $\calS$ is ex-ante single-peaked; and
    \item {\small $\E\brac{|q-s| \bmid (q,s) \in \calD} \le
    \E\brac{|q-\ulg_T(s)| \bmid (q,s) \in \calD}$.}
\end{enumerate}
Then, the first implication of \Cref{cor:weak-symm} holds. That is, uniform letter grading with $T$ grades is at least as good as numerical scoring if $\alpha_m > \alpha_d$, the converse holds if $\alpha_m < \alpha_d$, and the two have equal performance if $\alpha_m = \alpha_d$.
\end{theorem}

Before diving into the proof, let us make a remark regarding the third technical condition in \Cref{thm:weak-symm-impr}. The technical condition states that, averaged over all such pairs, the true quality is closer to the score than to the midpoint of the interval that they both belong to. Later, we show that this condition is satisfied in two natural cases. Intuitively, if the score distribution is sufficiently concentrated near the true quality, the expected distance between the score and the true quality will be sufficiently small, satisfying the condition.
Let us now turn to the proof of \Cref{thm:weak-symm-impr}.

\begin{proof}
Given \Cref{thm:weak-symm}, we simply need to show that $\E\brac{|q-s|} \le \E\brac{|q-\ulg_T(s)|}$. We already assume that this holds conditioned on $(q,s) \in \calD$. Hence, we only need to show that it also holds conditioned on $(q,s) \notin \calD$. We show this given the additional single-peakedness property. In fact, we show that conditioned on $(q,s) \notin \calD$, the desired equation actually holds for every $q \in [0,1]$, and, thus, in expectation over $q \sim \calQ$ too. Fix any $q \in [0,1]$. Note that
\allowdisplaybreaks
\begin{align*}
&\E\brac{|q-s| \bmid (q,s) \notin \calD}\\
&\quad= \Pr\brac{\ulg_T(s) < \ulg_T(q) \bmid (q,s) \notin \calD} 
\cdot \E\brac{q-s \bmid \ulg_T(s) < \ulg_T(q)} \\
&\quad\quad+ \Pr\brac{\ulg_T(s) > \ulg_T(q) \bmid (q,s) \notin \calD} \cdot \E\brac{s-q \bmid \ulg_T(s) > \ulg_T(q)}\\
&\quad\le \Pr\brac{\ulg_T(s) < \ulg_T(q) \bmid (q,s) \notin \calD} \cdot \E\brac{q-\ulg_T(s) \bmid \ulg_T(s) < \ulg_T(q)} \\
&\quad\quad+ \Pr\brac{\ulg_T(s) > \ulg_T(q) \bmid (q,s) \notin \calD} \cdot \E\brac{\ulg_T(s)-q \bmid \ulg_T(s) > \ulg_T(q)}\\
&\quad=\E\brac{|q-\ulg_T(s)| \bmid (q,s) \notin \calD},
\end{align*}
where the first transition holds because 
\[
[0,1]^2 \setminus \calD = \set{(q,s) : \ulg_T(s) < \ulg_T(q)} \cup \set{(q,s) : \ulg_T(s) > \ulg_T(q)},
\]
and the second transition holds due to linearity of expectation and because the single-peakedness assumption implies
\begin{align*}
 &\E\brac{s \bmid \ulg_T(s) < \ulg_T(q)}\ge \E\brac{\ulg_T(s) \bmid \ulg_T(s) < \ulg_T(q)},  \\
 \text{(and)}~~ &\E\brac{s \bmid \ulg_T(s) > \ulg_T(q)} \le \E\brac{\ulg_T(s) \bmid \ulg_T(s) > \ulg_T(q)}. 
\end{align*}
This completes the proof.
\end{proof}

In \Cref{thm:weak-symm-impr}, we argued that single-peakedness of $\calS$ establishes the desired inequality of $\E\brac{|q-s|} \le \E\brac{|q-\ulg_T(s)|}$ at least conditioned on $(q,s) \notin \calD$, leaving only the case of $(q,s) \in \calD$, which was stated as an assumption in. 
Next, we show that if the true quality prior $\calQ$ is uniform over $[0,1]$, and it satisfies two natural assumptions then the desired inequality also holds conditioned on $(q,s) \in \calD$. 

\begin{definition}[Ex-Post Single-Peaked Score Model]
We say that the score model $\calS$ is \emph{ex-post single-peaked} if, for every $s \in [0,1]$, $f_{\calS}(s;\cdot)$ is single-peaked with the peak at $s$, i.e., $f_{\calS}(s;q) \le f_{\calS}(s;q')$ for all $s \le q' \le q$ and $q \le q' \le s$.
\end{definition}

\begin{definition}[Probabilistic Single-Dipped Score Model]
We say that the score model $\calS$ is \emph{probabilistic single-dipped} if, for every $x \in [0,1]$, $\Pr\brac{s \in [q,x] \cup [x,q] \bmid q}$ (let us call this $p(x,q)$) is single-dipped in $q$ with the dip at $q=x$, i.e., $p(x,q) \le p(x,q')$ for all $x \le q' \le q$ and $q \le q' \le x$. 
\end{definition}

Before we state the next theorem, we further partition $\calD$ into two sub-spaces, $\calD^{\same}$ and $\calD^{\opp}$, such that $\calD^{\same}$ contains the set of all pairs of true qualities and scores such that either both are at most or both are at least the midpoint of their common letter grade interval, i.e.
\begin{align*}
 &\calD^{\same} = \{(q,s) : q,s \le \ulg_T(q)=\ulg_T(s) \lor q,s \ge \ulg_T(q)=\ulg_T(s) \}
\end{align*}
and  $\calD^{\opp} =  \calD \setminus \calD^{\same}$. For example,  when $T=10$, $(q=0.54,s=0.51) \in \calD^{\same}$, but $(q=0.54,s'=0.56) \in \calD^{\opp}$. We are now ready to state the result.
\begin{theorem}\label{thm:weak-symm-impr-2}
Fix arbitrary $T \in \bbN$. Assume the following regarding the true quality prior $\calQ$ and the score model $\calS$.
\begin{enumerate}
    \item\label{thm3:uniform} $\calQ$ is uniform over $[0,1]$;
    \item\label{thm3:joint-symm} $\calQ$ and $\calS$ are jointly symmetric;
    \item\label{thm3:peak-dip} $\calS$ is ex-ante and ex-post  single-peaked,  and probabilistic single-dipped; and
    \item\label{thm3:constant} $\Pr\brac{(q,s) \in \calD^{\same}} \geq 2(\gamma+1) \cdot \Pr\brac{(q,s) \in \calD^{\opp}}$, where $\gamma = \max_{a,b \in [0,1]} \frac{f_{\calS}(a;b)}{f_{\calS}(b;a)}$.
\end{enumerate}
Then, the first implication of \Cref{cor:weak-symm} holds. That is, uniform letter grading with $T$ grades is at least as good as numerical scoring if $\alpha_m > \alpha_d$, the converse holds if $\alpha_m < \alpha_d$, and the two have equal performance if $\alpha_m = \alpha_d$.
\end{theorem}
Let us first understand the assumptions in \Cref{thm:weak-symm-impr-2}. A natural choice of $\calS$ under which Assumptions 3 and 4 in \Cref{thm:weak-symm-impr-2} are satisfied is when $\calS(q)$ is a symmetric distribution around $q$, i.e., the noise in the score follows a symmetric zero-mean distribution. Further, for such a score model, we have $\gamma = 1$, so Assumption 4 becomes $\Pr[(q,s) \in \calD^{\same}] \geq 4 \cdot \Pr[(q,s) \in \calD^{\opp}]$. More general, from the definitions of $\calD^{\same}$ and $\calD^{\opp}$, when the variance of the score distribution is sufficiently small, we can expect $\Pr[(q,s) \in \calD^{\same}]$ to be much higher than $\Pr[(q,s) \in \calD^{\opp}]$. For further intuition, see \Cref{fig:intuition} in \Cref{app:intuition}.

\begin{proof}[Proof of \Cref{thm:weak-symm-impr-2}]
Given \Cref{thm:weak-symm-impr}, we only need to show that 
\begin{align}\label{eqn:4-to-prove}
\begin{split}
&\E\brac{|q-s| - |q-\ulg_T(s)| \bmid (q,s) \in \calD}\\
&\qquad= \Pr[(q,s) \in \calD^{\same} \bmid (q,s) \in \calD] \cdot \E\brac{|q-s| - |q-\ulg_T(s)| \bmid (q,s) \in \calD^{\same}}\\
&\qquad\qquad + \Pr[(q,s) \in \calD^{\opp} \bmid (q,s) \in \calD] \cdot \E\brac{|q-s| - |q-\ulg_T(s)| \bmid (q,s) \in \calD^{\opp}}\\
&\qquad\le 0.
\end{split}
\end{align}

Let us analyze the expected value of $|q-s| - |q-\ulg_T(s)|$ conditioned on both $(q,s) \in \calD^{\same}$ and $(q,s) \in \calD^{\opp}$ separately.

\paragraph{Analyzing $\boldsymbol{\calD^{\same}}$.} For $k \in \set{0,1,\ldots,T-1}$, define $\ell(k) = k\Delta$, $m(k) = (k+1/2)\Delta$, and $h(k) = (k+1)\Delta$. These are respectively the lower end, midpoint, and upper end of the $k$-th grade interval under $\ulg_T$. Note that 
\begin{align*}
\calD^{\same} = \Big\{(q,s): &(\ell(k) \le q \le s \le m(k)) \lor (\ell(k) \le s \le q \le m(k)) \lor \\
&(m(k) \le q \le s < h(k)) \lor (m(k) \le s \le q < h(k)), \ k \in \set{0,1,\ldots,T-1} \Big\}.
\end{align*}

Fix an arbitrary $k \in \set{0,1,\ldots,T-1}$; write $\ell$, $m$, and $h$ while omitting the fixed $k$ in the argument; and let us analyze the desired expression $|q-s| - |q-\ulg_T(s)|$ conditioned on each of the four cases for this fixed $k$ separately. We will derive bounds that will hold regardless of the value of $k$, and, therefore, also conditional on $(q,s) \in \calD^{\same}$ (i.e., aggregated across all $k$). Note that in each case, we have $\ulg_T(q)=\ulg_T(s)=m$.

\begin{enumerate}
    \item $\ell \le q \le s \le m$. In this case, $|q-s| - |q-\ulg_T(s)| = s-m$. Note that
    \begin{align*}
        \E\brac{s-m \mid \ell \le q \le s \le m} &= \frac{\int_{q=\ell}^m \int_{s=q}^m f_{\calQ}(q) \cdot f_{\calS}(s;q) \cdot (s-m) \dif s \dif q}{\Pr[\ell \le q \le s \le m]}\\
        &= \frac{\int_{q=\ell}^m 1 \cdot \E\brac{s-m \bmid q, s \in [q,m]} \cdot \Pr\brac{s \in [q,m] \bmid q} \dif q}{\int_{q=\ell}^m 1 \cdot \Pr\brac{s \in [q,m] \bmid q} \dif q}\\
        &\le \frac{-\int_{q=\ell}^m \left(\frac{m-q}{2}\right) \cdot \Pr\brac{s \in [q,m] \bmid q} \dif q}{\int_{q=\ell}^m 1 \cdot \Pr\brac{s \in [q,m] \bmid q} \dif q}\\
        &\le \frac{-\frac{1}{m-\ell} \cdot \left(\int_{q=\ell}^m \frac{m-q}{2} \dif q\right) \cdot \left(\int_{q=\ell}^m \Pr\brac{s \in [q,m] \bmid q} \dif q\right)}{\int_{q=\ell}^m 1 \cdot \Pr\brac{s \in [q,m] \bmid q} \dif q}\\
        &=-\frac{1}{(\Delta/2)} \int_{r=0}^{\frac{\Delta}{2}} \frac{r}{2} \dif r =-\frac{\Delta}{8}.
    \end{align*}
    Here, the third transition holds because conditioned on a given value of $q$ and on $s \in [q,m]$, the distribution of $s \in [q,m]$ is single-peaked with peak at $q$ (Assumption~\ref{thm3:peak-dip}). Hence, $\E[s | q,s \in [q,m]] \le (q+m)/2$. The fourth transition is the integral Chebyshev inequality (\Cref{lem:cheb} in \Cref{app:lemmas}), which holds because both $(m-q)/2$ and $\Pr\brac{s \in [q,m] \bmid q}$ are non-negative, non-increasing functions of $q$ in $[\ell,m]$ (Assumption~\ref{thm3:peak-dip}). 
    
    \item $\ell \le s \le q \le m$. In this case, $|q-s| - |q-\ulg_T(s)| = 2q-m-s$. Note that
    \begin{align*}
        &\E\brac{2q-m-s \mid \ell \le s \le q \le m} \\
        &\qquad= \frac{\int_{q=\ell}^m \int_{s=\ell}^q f_{\calQ\times\calS}(q,s) \cdot (2q-m-s) \dif s \dif q}{\Pr[\ell \le s \le q \le m]}\\
        &\qquad= \frac{\int_{s=\ell}^m f_{\calS}(s) \E\brac{2q-m-s \bmid s, q \in [s,m]} \cdot \Pr\brac{q \in [s,m] \bmid s} \dif s}{\Pr[\ell \le s \le q \le m]}.
    \end{align*}
    Here, we use $f_{\calQ\times\calS}(q,s) = f_{\calQ}(q) \cdot f_{\calS}(s;q)$ to denote the joint probability density of $q$ and $s$, and $f_{\calS}(s) = \int_{q=0}^1 f_{\calQ}(q) f_{\calS}(s;q) \dif q$ to denote the marginal probability density of $s$.
    
    We argue that $\E\brac{2q-m-s \bmid s, q \in [s,m]} \le 0$. Intuitively, this is because the posterior distribution of $q \in [s,m]$ conditioned on a fixed value of $s$ and on $q \in [s,m]$ is single-peaked with peak at $s$ by Assumptions~\ref{thm3:uniform} and~\ref{thm3:peak-dip}. Hence, $\E\brac{q \bmid s, q \in [s,m]} \le (s+m)/2$. Formally, this can be viewed as
    \begin{align*}
    \E\brac{2q-m-s \bmid s, q \in [s,m]} &= \frac{\int_{q=s}^m f(q;s) (2q-m-s) \dif q}{\int_{q=s}^m f(q;s) \dif q}\\
    &\le \frac{\frac{1}{m-s} \cdot \left( \int_{q=s}^m f(q;s) \dif q \right) \cdot \left( \int_{q=s}^m (2q-m-s) \dif q \right)}{\int_{q=s}^m f(q;s) \dif q} = 0,
    \end{align*}
    where $f(q;s) = \frac{f_{\calQ}(q) \cdot f_{\calS}(s;q)}{f_{\calS}(s)}$ denotes the probability density of true quality being $q$ conditioned on the score being $s$; the second transition is the integral Chebyshev inequality (\Cref{lem:cheb} in \Cref{app:lemmas}), which holds because $f(q;s)$ is a non-increasing function of $q$ whereas $2q-m-s$ is a non-decreasing function of $q$;\footnote{To see why $f(q;s) = \frac{f_{\calQ}(q) \cdot f_{\calS}(s;q)}{f_{\calS}(s)}$ is non-increasing in $q$, note that the denominator does not depend on $q$ whereas the numerator is equal to $f_{\calS}(s;q)$ (Assumption~\ref{thm3:uniform}), which is non-increasing in $q$ (Assumption~\ref{thm3:peak-dip}).}\textsuperscript{\normalfont,}\footnote{Technically, integral Chebyshev inequality requires non-negative functions, and $2q-(m+s)$ can be negative when $q < (m+s)/2$. However, one can equivalently separate out the $-(m+s)$ term, apply the integral Chebyshev inequality to $2q$, and recombine with the $-(m+s)$ term to achieve the same conclusion.} and the final transition holds because the second integral in the numerator is $0$. 
    
    \item $m \le q \le s < h$. In this case, $|q-s| - |q-\ulg_T(s)| = m+s-2q$. Due to the same reasoning as in Case 2, we have that $\E\brac{m+s-2q \bmid m \le q \le s < h} \le 0$.
    
    \item $m \le s \le q < h$. In this case, $|q-s| - |q-\ulg_T(s)| = m-s$. Due to the same reasoning as in Case 1, we have that $\E\brac{m-s \bmid m \le s \le q < h} \le -\Delta/8$.
\end{enumerate}

Let $p_1,p_2,p_3,p_4$ respectively denote the total probabilities of the above four cases across all values of $k \in \set{0,1,\ldots,T-1}$, conditioned on $(q,s) \in \calD^{\same}$. Then, $p_1+p_2+p_3+p_4 = 1$. Because $f_{\calS}(a;b)/f_{\calS}(b;a) \le \gamma$ for all $a,b \in [0,1]$, it follows that $p_1 \ge p_2/\gamma$ and $p_4 \ge p_3/\gamma$. Hence, $p_1+p_4 \ge (p_2+p_3)/\gamma$. Using $p_1+p_2+p_3+p_4 = 1$, we get $p_1+p_4 \ge 1/(\gamma+1)$. 

Combining the analysis from the four cases above, we have 
\begin{equation}\label{eqn:4-dp}
    \E\brac{|q-s| - |q-\ulg_T(s)| \bmid (q,s) \in \calD^{\same}} \le -(p_1+p_4) \cdot \frac{\Delta}{8} \le -\frac{\Delta}{8(\gamma+1)}.
\end{equation}

\paragraph{Analyzing $\boldsymbol{\calD^{\opp}}$.} Note that 
\[
\calD^{\opp} = \cup_{k \in \set{0,1,\ldots,T-1}} \set{(q,s) : (\ell(k) \le q \le m(k) \le s \le h(k)) \lor (\ell(k) \le s \le m(k) \le q \le h(k))}.
\]

Fix an arbitrary $k \in \set{0,1,\ldots,T-1}$; as before, write $\ell$, $m$, and $h$ while omitting the fixed $k$ in the argument. Once again, we analyze the desired expression $|q-s| - |q-\ulg_T(s)|$ conditioned on each of the two cases in the above expansion of $\calD^{\opp}$ for this fixed $k$ separately. We will derive bounds that will hold regardless of the value of $k$, and, therefore, also conditional on $(q,s) \in \calD^{\opp}$ (i.e., aggregated across all $k$). Note that we still have $\ulg_T(q)=\ulg_T(s)=m$.

\begin{enumerate}
    \item $\ell \le q \le m \le s \le h$: In this case, we have $|q-s| - |q-\ulg_T(s)| = s-m$. Note that 
    \begin{equation}\label{eqn:4-dpp-1}
    \E\brac{s-m \bmid \ell \le q \le m \le s \le h} \le \Delta/4.
    \end{equation}
    This is because $s \in [m,m+\Delta/2]$ and, due to single-peakedness of the score model and $q \le m$, it is at most $m+\Delta/4$ in expectation. 

    \item $\ell \le s \le m \le q \le h$: In this case, we have $|q-s| - |q-\ulg_T(s)| = m-s$, and the same reasoning as above shows that 
    \begin{equation}\label{eqn:4-dpp-2}
    \E\brac{m-s \bmid \ell \le s \le m \le q \le h} \le \Delta/4.
    \end{equation}
\end{enumerate}

Combining \Cref{eqn:4-dpp-1,eqn:4-dpp-2} and aggregating over all $k \in \set{0,1,\ldots,T-1}$, we get that
\begin{equation}\label{eqn:4-dpp}
\E\brac{|q-s| - |q-\ulg_T(s)| \bmid (q,s) \in \calD^{\opp}} \le \Delta/4.
\end{equation}

Finally, combining \Cref{eqn:4-dp,eqn:4-dpp}, we have that 
\begin{align*}
&\E\brac{|q-s| - |q-\ulg_T(s)| \bmid (q,s) \in \calD}\\
&\quad\le \Pr\brac{(q,s) \in \calD^{\same} \bmid (q,s) \in \calD} \cdot \left(-\frac{\Delta}{8(\gamma+1)}\right) + \Pr\brac{(q,s) \in \calD^{\opp} \bmid (q,s) \in \calD} \cdot \frac{\Delta}{4} \le 0,
\end{align*}
where the final transition holds because $\Pr\brac{(q,s) \in \calD^{\same}} \ge 2(\gamma+1) \cdot \Pr\brac{(q,s) \in \calD^{\opp}}$ (Assumption~\ref{thm3:constant}), which is equivalent to 
\[
\Pr\brac{(q,s) \in \calD^{\same} \bmid (q,s) \in \calD} \ge 2(\gamma+1) \cdot \Pr\brac{(q,s) \in \calD^{\same} \bmid (q,s) \in \calD}.
\]
This completes the proof.
\end{proof}

Ex-ante single-peakedness, ex-post single-peakedness, and probabilistic single-dippedness can be subsumed into a single property that captures a stronger form of symmetry, in which the noise in the score is symmetric and zero-mean. 
\begin{definition}[Strongly Symmetric Score Model]
We say that the score model $\calS$ is strongly symmetric if $f_{\calS}(s;q) = \ell(|s-q|)$ for some non-increasing function $\ell:\bbR_{\ge 0} \to \bbR_{\ge 0}$.
\end{definition}

Under a strongly symmetric score model, we have $\gamma = 1$ in Assumption~\ref{thm3:constant} of \Cref{thm:weak-symm-impr-2}, which means a constant of $2(\gamma+1) = 4$ would be needed. However, using different techniques, we can show that even a constant of $3$ suffices to obtain the same result under strong symmetry. This broadens the scope to include less concentrated score models.  

\begin{theorem}\label{clm:threetimesbucketing}
Fix arbitrary $T \in \bbN$. Let $\calD$, $\calD^{\same}$, and $\calD^{\opp}$ be defined as in \Cref{thm:weak-symm-impr-2}. Assume the following regarding the true quality prior $\calQ$ and the score model $\calS$.
\begin{enumerate}
    \item\label{thm4:uniform} $\calQ$ is uniform over $[0,1]$;
    \item\label{thm4:strong-symm} $\calS$ is strongly symmetric; and
    \item\label{thm4:constant} $\Pr\brac{(q,s) \in \calD^{\same}} \geq 3 \cdot \Pr\brac{(q,s) \in \calD^{\opp}}$.
\end{enumerate}
Then, the first implication of \Cref{cor:weak-symm} holds. That is, uniform letter grading with $T$ grades is at least as good as numerical scoring if $\alpha_m > \alpha_d$, the converse holds if $\alpha_m < \alpha_d$, and the two have equal performance if $\alpha_m = \alpha_d$.
\end{theorem}
\begin{proof}
As in the proof of \Cref{thm:weak-symm-impr-2}, note that given \Cref{thm:weak-symm-impr}, we only need to prove 
\begin{align}\label{eqn:threetimesmain}
\begin{split}
&\E\brac{|q-s| - |q-\ulg_T(s)| \bmid (q,s) \in \calD}\\
&\qquad= \Pr[(q,s) \in \calD^{\same} \bmid (q,s) \in \calD] \cdot \E\brac{|q-s| - |q-\ulg_T(s)| \bmid (q,s) \in \calD^{\same}}\\
&\qquad\qquad + \Pr[(q,s) \in \calD^{\opp} \bmid (q,s) \in \calD] \cdot \E\brac{|q-s| - |q-\ulg_T(s)| \bmid (q,s) \in \calD^{\opp}}\\
&\qquad\le 0.
\end{split}
\end{align}

In the proof of \Cref{thm:weak-symm-impr-2}, we analyzed the expected value of $|q-s| - |q-\ulg_T(s)|$ conditioned on both $(q,s) \in \calD^{\same}$ and $(q,s) \in \calD^{\opp}$ separately: the former was shown to be at most $-\frac{\Delta}{8(\gamma+1)}$ whereas the latter was shown to be at most $\frac{\Delta}{4}$, yielding the desired \Cref{eqn:threetimesmain} when $\Pr[(q,s) \in \calD^{\same}] \ge 2(\gamma+1) \cdot \Pr[(q,s) \in \calD^{\opp}]$. 

With strong symmetry (Assumption~\ref{thm4:strong-symm}), we improve the former upper bound to $-\frac{\Delta}{12}$, which improves the sufficient condition to $\Pr[(q,s) \in \calD^{\same}] \ge 3 \cdot \Pr[(q,s) \in \calD^{\opp}]$. That is, our goal is to prove
\[
\E\brac{|q-s| - |q-\ulg_T(s)| \bmid (q,s) \in \calD^{\same}} \le -\frac{\Delta}{12}.
\]

Note that $\calD^{\same} = \cup_{k \in \set{0,1,\ldots,T-1}} \calD^{\same}_k$, where $\calD^{\same}_k = \calD^{\same} \cap [k\Delta,(k+1)\Delta)^2$. We show that $\E\brac{|q-s| - |q-\ulg_T(s)| \bmid (q,s) \in \calD^{\same}_k} \le -\frac{\Delta}{12}$ for all $k \in \set{0,1,\ldots,T-1}$, which implies the desired result. Fix any $k \in \set{0,1,\ldots,T-1}$, and write $\ell = k\Delta$, $m = (k+1/2)\Delta$, and $h = (k+1)\Delta$.

Let us further partition $\calD^{\same}_k$ as $\calD^{\same}_{k,\low} \cup \calD^{\same}_{k,\high}$, where $\calD^{\same}_{k,\low} = \set{(q,s) : \ell \le q,s < m}$ (both the true quality and the score are lower than the midpoint) and $\calD^{\same}_{k,\high} = \set{(q,s) : m \le q,s < h}$ (both the true quality and the score are at least as high as the midpoint). Crucially, we note that 
\[
\E\brac{|q-s| - |q-m| \bmid (q,s) \in \calD^{\same}_{k,\low}} = \E\brac{|q-s| - |q-m| \bmid (q,s) \in \calD^{\same}_{k,\high}}.
\]
This is because the transformation $(q,s) \to (q',s')$, where $q' = m+(m-q)$ and $s' = m+(m-s)$, is a bijection mapping each point $(q,s) \in \calD^{\same}_{k,\low}$ to a point $(q',s') \in \calD^{\same}_{k,\high}$ with $|q-s| - |q-m| = |q'-s'| - |q'-m|$ as well as $f_{\calQ \times \calS}(q,s) = f_{\calQ \times \calS}(q',s')$; the last observation relies on $\calQ$ being a uniform distribution (Assumption~\ref{thm4:uniform}) and $\calS$ being strongly symmetric (Assumption~\ref{thm4:strong-symm}). 

Hence, it is sufficient to show that 
\[
\E\brac{|q-s| - |q-m| \bmid (q,s) \in \calD^{\same}_{k,\low}} \le -\frac{\Delta}{12}.
\]

Next, we further partition $\calD^{\same}_{k,\low}$ as $\calD^{\same}_{k,\low,\inc} \cup \calD^{\same}_{k,\low,\dec}$, where $\calD^{\same}_{k,\low,\inc} = \set{(q,s) : \ell \le q \le s < m}$ (the score is at least as much as the true quality) and $\calD^{\same}_{k,\low,\dec} = \set{(q,s) : \ell \le s \le q < m}$ (the score is at most as much as the true quality). Note that 
\begin{align*}
&\E\brac{|q-s| - |q-m| \bmid (q,s) \in \calD^{\same}_{k,\low}} \\
&\quad = \Pr\brac{(q,s) \in \calD^{\same}_{k,\low,\inc} \bmid (q,s) \in \calD^{\same}_{k,\low}} \cdot \E\brac{|q-s| - |q-m| \bmid (q,s) \in \calD^{\same}_{k,\low,\inc}}\\
&\quad +\Pr\brac{(q,s) \in \calD^{\same}_{k,\low,\dec} \bmid (q,s) \in \calD^{\same}_{k,\low}} \cdot \E\brac{|q-s| - |q-m| \bmid (q,s) \in \calD^{\same}_{k,\low,\dec}}.
\end{align*}

First, we argue that 
\[
\Pr\brac{(q,s) \in \calD^{\same}_{k,\low,\inc} \bmid (q,s) \in \calD^{\same}_{k,\low}} = \Pr\brac{(q,s) \in \calD^{\same}_{k,\low,\dec} \bmid (q,s) \in \calD^{\same}_{k,\low}} = \frac{1}{2}.
\]
This follows by noting the bijection from $\calD^{\same}_{k,\low,\inc}$ to $\calD^{\same}_{k,\low,\dec}$ given by $(q,s) \to (q',s')$, where $q' = m-(q-\ell)$ and $s' = m-(s-\ell)$; due to strong symmetry of $\calS$ and $|q-s|=|q'-s'|$, we have $f_{\calQ\times\calS}(q,s)=f_{\calQ\times\calS}(q',s')$. 

Next, recall that in the proof of \Cref{thm:weak-symm-impr-2} (Case 2 in the analysis of $\calD^{\same}$), we had already argued 
\[
\E\brac{|q-s| - |q-m| \bmid (q,s) \in \calD^{\same}_{k,\low,\dec}} = \E[2q-s-m | \ell \le s \le q < m] \le 0.
\]
Hence, we have
\[
\E\brac{|q-s| - |q-m| \bmid (q,s) \in \calD^{\same}_{k,\low}} \le \frac{1}{2} \cdot \E\brac{|q-s| - |q-m| \bmid (q,s) \in \calD^{\same}_{k,\low,\inc}},
\]
which means it is sufficient to argue
\[
\E\brac{|q-s| - |q-m| \bmid (q,s) \in \calD^{\same}_{k,\low,\inc}} = \E\brac{s-m \bmid \ell \le q \le s < m} \le -\frac{\Delta}{6}.
\]

Note that 
\begin{align*}
&\E\brac{s-m \bmid \ell \le q \le s < m} \\
&\quad= -\frac{\int_{q=\ell}^m \int_{s=q}^m (m-s) f_{\calS}(s;q) \dif s \dif q}{\Pr[\ell \le q \le s < m]} &&\text{($\calQ$ is uniform)}\\
&\quad\le -\frac{\int_{q=\ell}^m \frac{1}{m-q} \left(\int_{s=q}^m (m-s) \dif s\right) \cdot \left(\int_{s=q}^m f_{\calS}(s;q) \dif s\right) \dif q}{\Pr[\ell \le q \le s < m]} &&\text{(\Cref{lem:cheb} in \Cref{app:lemmas})}\\
&\quad= -\frac{1}{2} \frac{\int_{q=\ell}^m (m-q) \Pr[s \in [q,m]] \dif q}{\Pr[\ell \le q \le s < m]}\\
&\quad\le -\frac{1}{2} \frac{\frac{2(m-\ell)}{3} \cdot \int_{q=\ell}^m \Pr[s \in [q,m]] \dif q}{\Pr[\ell \le q \le s < m]} &&\text{(\Cref{lem_fd_last} in \Cref{app:lemmas})}\\
&\quad= -\frac{1}{2} \frac{\frac{2(m-\ell)}{3} \cdot \Pr[\ell \le q \le s < m]}{\Pr[\ell \le q \le s < m]} = -\frac{\Delta}{6},
\end{align*}
as needed. Here, in the application of \Cref{lem_fd_last} from \Cref{app:lemmas} in the fourth transition, we use the fact that $g(q) = \Pr[s \in [q,m]] = \int_{s=q}^m f_{\calS}(s;q) \dif s$ is a concave function and $g(m) = 0$. To see concavity, note that strong symmetry of $\calS$ means that there is a distribution with probability density $z$ such that $f_{\calS}(s;q) = z(s-q)$. Then, $g(q) = \int_{s=q}^m z(s-q) \dif s = \int_{x=0}^{m-q} z(x) \dif x$. Hence, $g'(q) = -z(m-q)$ and $g''(q) = z'(m-q)$. Due to the single-peakedness of $\calS$, we have that $z'(x) \le 0$ for all $x \ge 0$, so $g''(q) \le 0$, which proves concavity of $g$. 
\end{proof}

We remark that in the proof of \Cref{clm:threetimesbucketing}, we only really need strong symmetry for pairs of true qualities and scores that belong to the same letter grade interval. 

\section{Experiments}\label{sec:experiments}
\begin{figure*}[t]
    \centering
 \begin{subfigure}[t]{0.47\textwidth}
        \includegraphics[width=\textwidth]{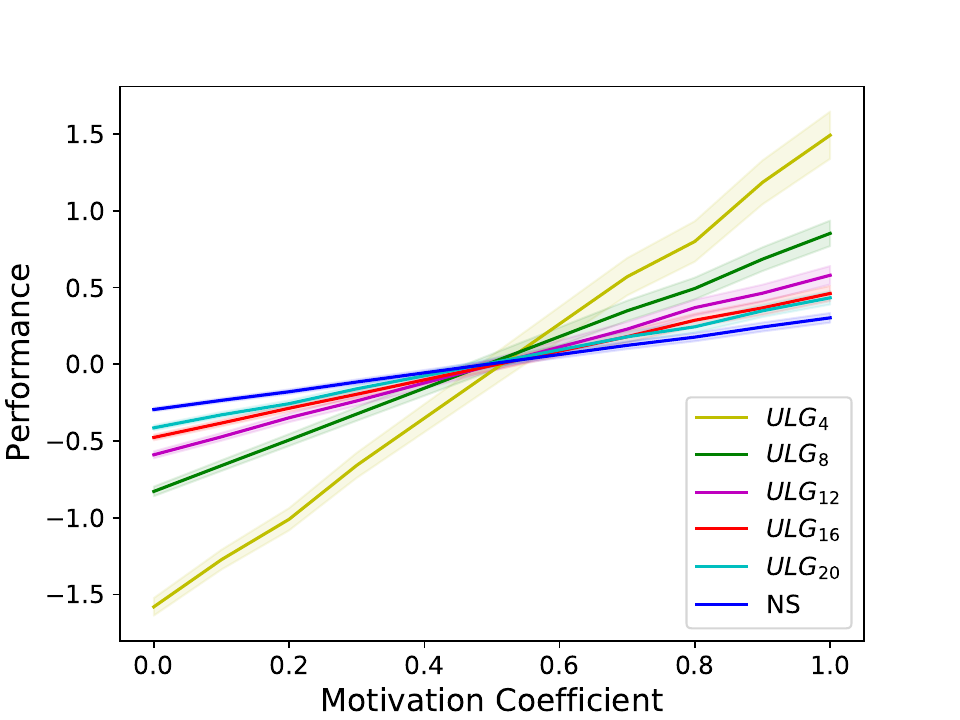}
        \caption{\textsc{$r=2$}}
        \label{subfig:mot-final-2}
    \end{subfigure}\hspace{0.05\textwidth}%
      \begin{subfigure}[t]{0.47\textwidth}
        \includegraphics[width=\textwidth]{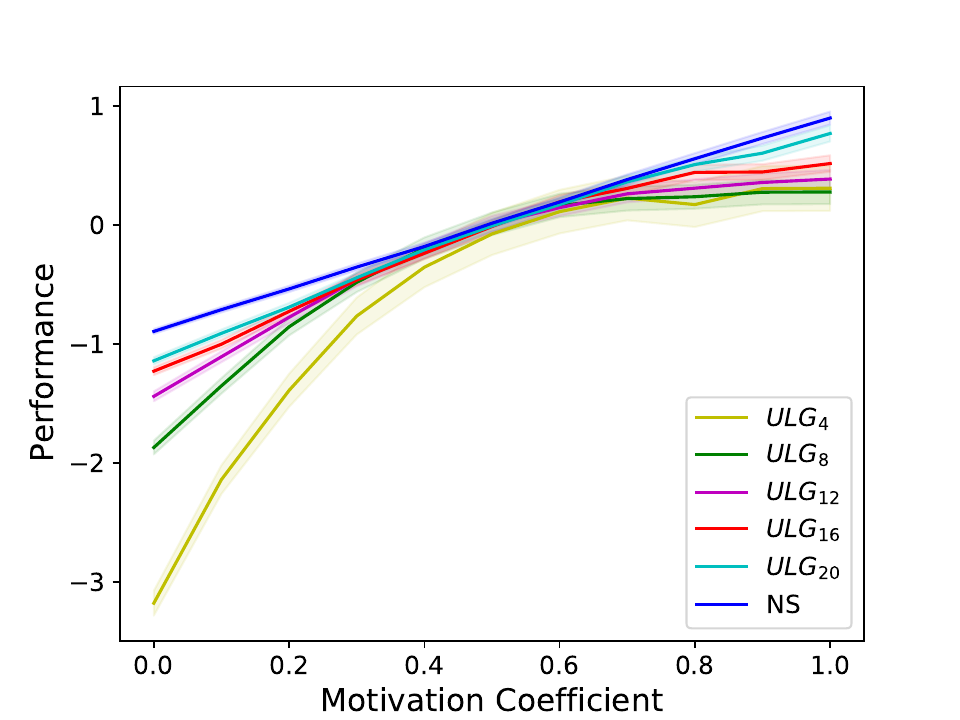}
        \caption{\textsc{$r=4$}}
        \label{subfig:mot-final-4}
    \end{subfigure}
    \begin{subfigure}[t]{0.47\textwidth}
        \includegraphics[width=\textwidth]{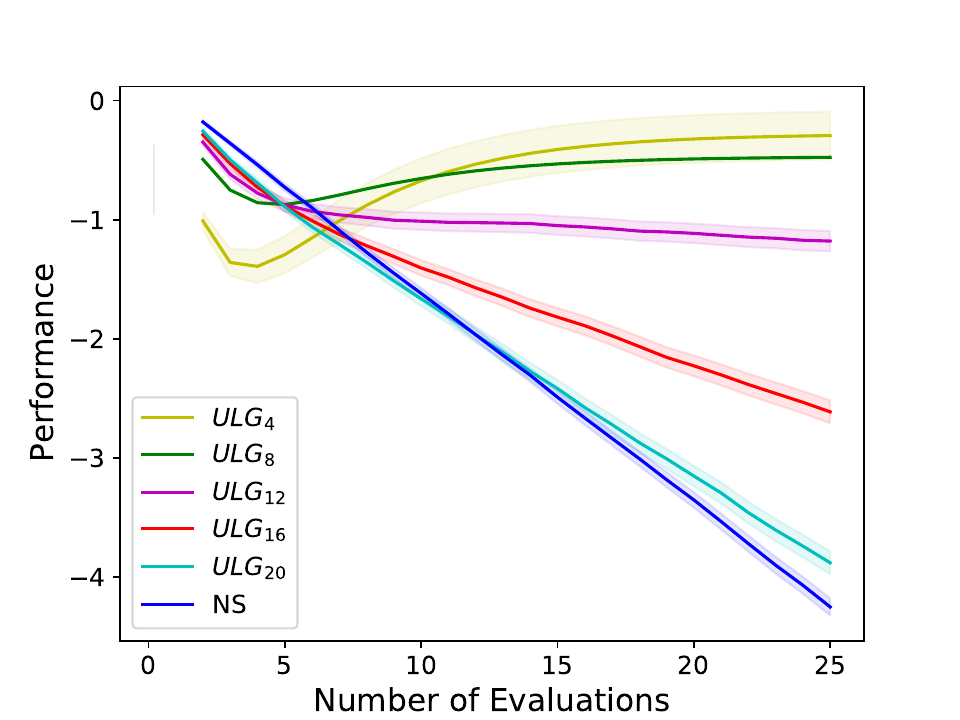}
        \caption{\textsc{$\alpha_m=0.2$}}
        \label{subfig:eval-final-0.2}
    \end{subfigure}\hspace{0.05\textwidth}%
    \begin{subfigure}[t]{0.47\textwidth}
        \includegraphics[width=\textwidth]{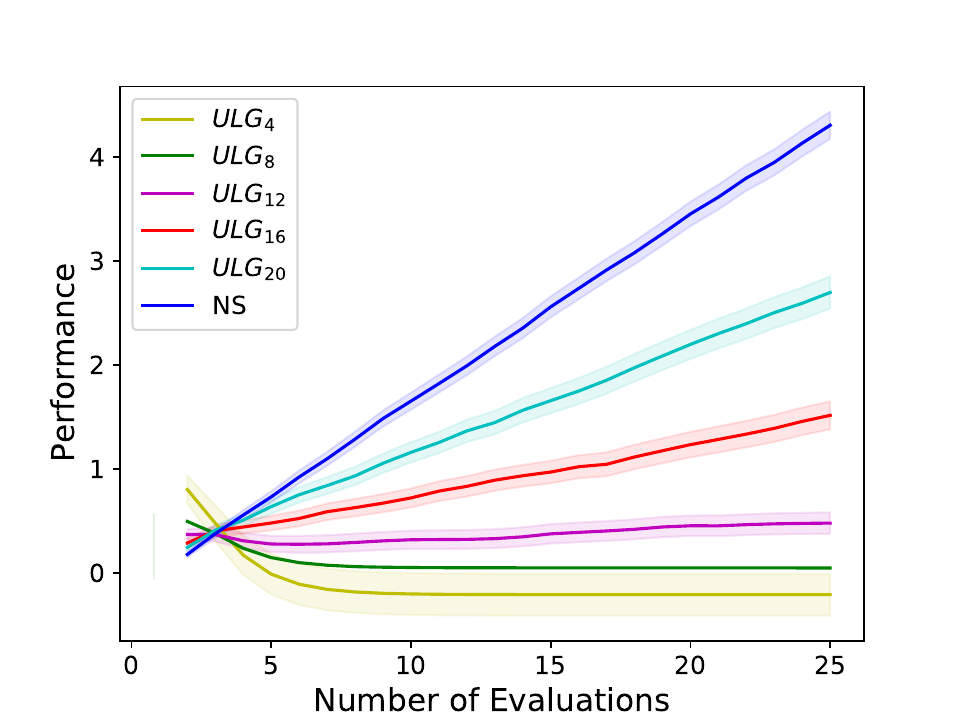}
        \caption{$\alpha_m=0.8$}
        \label{subfig:eval-final-0.8}
    \end{subfigure}
    \caption{Performance of numerical scoring and different uniform letter grading schemes, with $\mu=65$, $\sigma=12$,  $\gamma=1.5$ and $\alpha_d=0.5$ over different motivation coefficients (top) and number of evaluations (bottom). $95\%$ confidence intervals are shown.}
       \label{fig:final-score}
\end{figure*}

In the previous section, we proved that when $\calQ$ is uniformly distributed and the variance of the score model is small, we can conclude that the first implication of \Cref{cor:weak-symm} holds. In this section, we empirically compare numerical scoring and uniform letter grading while relaxing these assumptions. 

First, it is widely believed that students' true qualities, at least in large classes, are normally distributed based on the evidence that ``exam scores tend to be normally distributed for well-constructed, norm-referenced, multiple choice tests''~\cite{wedell1989student}. Hence, we empirically study the case where $\calQ$ is normally distributed, truncated to $[0,1]$. We also consider cases where the score is not necessarily concentrated around the true quality. Finally, our  analysis was limited to two evaluations; in our experiments, we also consider more than two evaluations. When a student participates in $r$ sequential evaluations, after each evaluation the student compares her ``current'' true quality to the obtained grade, and experiences (de)motivation that affects her effective true quality in the next evaluation. Formally, for $j \in [r]$, let $q_j$ and $s_j$ denote her effective true quality and score in evaluation $j$, respectively. Then, $s_j \sim \calS(q_j)$ for each $j \in [r]$, and for $j \in [r-1]$, we have:
\begin{align*}
    q_{j+1} = \begin{cases}
       q_{j}+ \alpha_m \cdot (\bucket(s_{j})-q_{j}), &\text{if $\bucket(s_{j}) \ge q_{j}$},\\
        q_{j} - \alpha_d \cdot (q_{j}-\bucket(s_{j})), &\text{if $\bucket(s_{j}) < q_{j}$}.
        \end{cases}
\end{align*}

We measure the performance of a grading scheme by comparing the final true quality, $q_r$, to the initial true quality $q_1$, which extends the performance measure introduced in preliminaries for two evaluations: 
\begin{align*}
     \perf_F(\bucket) \triangleq \E_{q \sim \calQ, s \sim \calS(q)}[ q_r-q_1].
\end{align*}

\smallskip\noindent\textbf{Data generation.} For all the simulations, we compare numerical scoring ($\ns$) to uniform letter grading ($\ulg_{T}$) with $T\in \{4,8,12,16,20\}$ grades. We scale the interval of grades to $[0,100]$ to resemble percentage grades. We simulate $n=5000$ students (average results are plotted with $95\%$ confidence intervals), where the initial true quality $q_1$ of each student is drawn i.i.d. from a truncated normal distribution capped to $[0,100]$, with the underlying normal distribution characterized by mean $\mu$ and standard deviation $\sigma$. Given a true quality $q$ in an evaluation, the score $s$ is drawn from another truncated normal distribution capped to $[0,100]$, with the underlying normal distribution characterized by mean $q$ and standard deviation $\gamma$. 

\smallskip\noindent\textbf{Results.} \Cref{fig:final-score} shows how the final quality improves (or degrades) with respect to the motivation coefficient (top) and the number of evaluations (bottom). In \Cref{subfig:mot-final-2}, the motivation coefficient takes values in $\{0,0.1\ldots,0.9,1\}$, the demotivation coefficient is set to $0.5$ and the number of evaluations is set to $r=2$. We see that when $\alpha_m<\alpha_d$, numerical scoring is better than any uniform letter grading (and uniform letter grading with more grades is better than uniform letter grading with fewer grades), whereas when $\alpha_m>\alpha_d$, the opposite is true. Hence, it seems that the first implication of \Cref{cor:weak-symm} continues to hold, even when the true quality is drawn from more realistic distributions. The comparison between uniform letter grading schemes with different numbers of grades is intuitive: uniform letter grading essentially converges to numerical scoring when $T$ goes to infinity, so larger $T$ should resemble numerical scoring more. The experiments show that this holds even with small values of $T$. 

Going beyond our theoretical analysis for $r=2$ evaluations, we consider the case where students participate in more than two evaluations. Surprisingly, as seen in \Cref{subfig:eval-final-0.2,subfig:eval-final-0.8}, the comparison between numerical scoring and uniform letter grading flips completely with large values of $r$: numerical scoring becomes \emph{worse} than uniform letter grading (and $\ulg_T$ becomes worse than $\ulg_{T'}$ for $T > T'$) when $\alpha_m < \alpha_d$, but \emph{better} when $\alpha_d < \alpha_m$. This shows that the choice of the grading scheme depends not only on the comparison between the strengths of motivational and demotivational effects ($\alpha_m$ vs $\alpha_d$) but also, crucially, on the number of evaluations $r$. When $\alpha_m < \alpha_d$, with fewer evaluations (e.g., courses with fewer tests or curricula with fewer semesters), use of numerical scoring may be recommended, whereas with many evaluations (e.g., courses with frequent tests or curricula with many semesters), use of uniform letter grading with fewer letters may be more appropriate. 

The transition between the regimes of few evaluations and many evaluations is even more surprising. As seen in \Cref{subfig:mot-final-4}, with $r=4$ evaluations, numerical scoring seems to outperform uniform letter grading schemes regardless of the comparison between $\alpha_m$ and $\alpha_d$. Hence, in general, it is always best to simulate different grading schemes under the model and the number of evaluations of interest in order to pick a suitable grading scheme. 

Finally, we observe that under numerical scoring, as the number of evaluations increases, the average student quality declines linearly when $\alpha_m < \alpha_d$ (\Cref{subfig:eval-final-0.2}) and improves linearly when $\alpha_m > \alpha_d$ (\Cref{subfig:eval-final-0.8}). This is expected because it can be shown that under numerical scoring, every evaluation changes the expected student quality by the same amount, which is proportional to $\alpha_m-\alpha_d$, leading to a linear decline or growth. In contrast, under uniform letter grading schemes with very few grades (small $T$), the average student quality seems to converge and remain stable as the number of evaluations increases, regardless of the comparison between $\alpha_m$ and $\alpha_d$. This can be explained due to the following stabilizing effect. Let $[\ell,h]$ be a letter grade interval and $m$ be its midpoint. Consider a student who starts with a true quality $q \in [\ell,h]$. The student is likely to receive a score $s$ in the same interval $[\ell,h]$ (so that $(q,s) \in \mathcal{D}$), and thus, a grade of $m$. This causes the true quality to update in a manner so that it gets closer to $m$ after which the student experiences very little motivation or demotivation due to receiving a grade that is almost equal to her true quality. Of course, the effect is more pronounced when $T$ is small, so letter grade intervals are large compared to the variance of the score model. 

We have presented only the most striking observations here. For additional experiments, see \Cref{app:more-figures}, where we notice that the first implication of \Cref{cor:weak-symm} continues to hold even when the score is not well-concentrated around the true quality. 

\section{Discussion}\label{sec:discussion}

Our work takes the first step towards proposing a statistical model of the  impact of letter grading schemes on student performance in sequential evaluations and using it to compare uniform letter grading schemes to numerical scoring. We view our work as a stepping stone and outline appealing extensions below.

\smallskip\noindent\textbf{Beyond midpoint grading.} In our model, we assume that if all the scores from an interval $[\ell,u]$ are mapped to the same grade, they are effectively mapped to the midpoint grade $(\ell+u)/2$. While this is a common method in practice 
of converting letter grades to percentages
~\cite{midpoint1,midpoint2}, other values within the range $[\ell,u]$ are also sometimes used~\cite{nonmidpoint}.   

\smallskip\noindent\textbf{Non-uniform letter grading.} It would be interesting to extend our analysis to non-uniform letter grading schemes.
More broadly,  how can our model be extended to incorporate truly non-numeric grades (e.g., A, B, etc.) without converting them to numeric grades somehow (e.g., 4, 3.7, etc.)?

\smallskip\noindent\textbf{Non-linear (de)motivation.}  Evidence from prospect theory suggests that motivational effects from positive outcomes are typically concave (diminishing rewards) while demotivational effects from negative outcomes are typically convex (increasing losses)~\cite{kahneman1979interpretation}. It would be interesting to study such nonlinear effects. 

\smallskip\noindent\textbf{Exploring implications to pedagogy and beyond.} There is a growing literature on optimizing design choices in AI-based learning systems, e.g., algorithmically deciding which explanations to show to students~\protect\cite{zavaleta2021using}. Our insights may inform the design of personalized grading schemes in such systems; they can adjust grade disclosure by learning over time whether students respond more strongly to motivation or demotivation.

More broadly, insights from our work can be explored in other multi-agent systems, such as contest design~\protect\cite{levy2017contest} and crowdsourcing~\protect\cite{han2021crowdsourcing}, where agents participate in rounds, and feedback from earlier rounds can influence the effort in subsequent rounds. For example, under the right conditions, \protect\Cref{thm:weak-symm-impr-2} may suggest a leaderboard design where teams are grouped into buckets (analogously to letter grading) and their exact performance is not revealed.

\bibliography{abb,bibliography}

\begin{thebibliography}{27}
\providecommand{\natexlab}[1]{#1}
\providecommand{\url}[1]{\texttt{#1}}
\expandafter\ifx\csname urlstyle\endcsname\relax
  \providecommand{\doi}[1]{doi: #1}\else
  \providecommand{\doi}{doi: \begingroup \urlstyle{rm}\Url}\fi

\bibitem[Baumeister et~al.(2001)Baumeister, Bratslavsky, Finkenauer, and
  Vohs]{baumeister2001bad}
R.~F. Baumeister, E.~Bratslavsky, C.~Finkenauer, and K.~D. Vohs.
\newblock Bad is stronger than good.
\newblock \emph{Review of general psychology}, 5\penalty0 (4):\penalty0
  323--370, 2001.

\bibitem[Bloodgood et~al.(2020)Bloodgood, Short, Jackson, and
  Martindale]{BSJ2020}
R.~A. Bloodgood, J.~G. Short, J.~M. Jackson, and J.~R. Martindale.
\newblock A change to pass/fail grading in the first two years at one medical
  school results in improved psychological well-being.
\newblock \emph{Economics of Education Review}, 84:\penalty0 655--662, 2020.

\bibitem[Brownback(2018)]{BROWNBACK2018113}
A.~Brownback.
\newblock A classroom experiment on effort allocation under relative grading.
\newblock \emph{Economics of Education Review}, 62:\penalty0 113--128, 2018.

\bibitem[Cameron and Pierce(1994)]{cameron1994reinforcement}
J.~Cameron and W.~D. Pierce.
\newblock Reinforcement, reward, and intrinsic motivation: A meta-analysis.
\newblock \emph{Review of Educational research}, 64\penalty0 (3):\penalty0
  363--423, 1994.

\bibitem[Christensen and Menzel(1998)]{christensen1998linear}
L.~J. Christensen and K.~E. Menzel.
\newblock The linear relationship between student reports of teacher immediacy
  behaviors and perceptions of state motivation, and of cognitive, affective,
  and behavioral learning.
\newblock \emph{Communication Education}, 47\penalty0 (1), 1998.

\bibitem[Coleman et~al.(1987)Coleman, Jussim, and Abraham]{coleman1987students}
L.~M. Coleman, L.~Jussim, and J.~Abraham.
\newblock Students' reactions to teachers' evaluations: The unique impact of
  negative feedback.
\newblock \emph{Journal of Applied Social Psychology}, 17\penalty0
  (12):\penalty0 1051--1070, 1987.

\bibitem[Czibor et~al.(2020)Czibor, Onderstal, Sloof, and {van
  Praag}]{CZIBOR2020}
E.~Czibor, S.~Onderstal, R.~Sloof, and M.~C. {van Praag}.
\newblock Does relative grading help male students? {E}vidence from a field
  experiment in the classroom.
\newblock \emph{Economics of Education Review}, 75:\penalty0 101953, 2020.

\bibitem[Deci et~al.(1999)Deci, Koestner, and Ryan]{deci1999meta}
E.~L. Deci, R.~Koestner, and R.~M. Ryan.
\newblock A meta-analytic review of experiments examining the effects of
  extrinsic rewards on intrinsic motivation.
\newblock \emph{Psychological bulletin}, 125\penalty0 (6):\penalty0 627, 1999.

\bibitem[Dev(1997)]{dev1997intrinsic}
P.~C. Dev.
\newblock Intrinsic motivation and academic achievement: What does their
  relationship imply for the classroom teacher?
\newblock \emph{Remedial and special education}, 18\penalty0 (1):\penalty0
  12--19, 1997.

\bibitem[Dubey and Geanakoplos(2010)]{dubey2010grading}
P.~Dubey and J.~Geanakoplos.
\newblock Grading exams: 100, 99, 98,… or {A}, {B}, {C}?
\newblock \emph{Games and Economic Behavior}, 69\penalty0 (1):\penalty0 72--94,
  2010.

\bibitem[Gray and Bunte(2022)]{gray2022effect}
T.~Gray and J.~Bunte.
\newblock The effect of grades on student performance: Evidence from a
  quasi-experiment.
\newblock \emph{College Teaching}, 70\penalty0 (1):\penalty0 15--28, 2022.

\bibitem[Guskey(2011)]{guskey2011five}
T.~R. Guskey.
\newblock Five obstacles to grading reform.
\newblock \emph{Educational Leadership}, 69\penalty0 (3):\penalty0 16, 2011.

\bibitem[Han et~al.(2020)Han, Tu, Yu, Wang, and
  Domeniconi]{han2021crowdsourcing}
G.~Han, J.~Tu, G.~Yu, J.~Wang, and C.~Domeniconi.
\newblock Crowdsourcing with multiple-source knowledge transfer.
\newblock In \emph{Proceedings of the 29th International Joint Conference on
  Artificial Intelligence (IJCAI)}, pages 2908--2914, 2020.

\bibitem[Kahneman and Tversky(1979)]{kahneman1979interpretation}
D.~Kahneman and A.~Tversky.
\newblock On the interpretation of intuitive probability: A reply to {J}onathan
  {C}ohen.
\newblock \emph{Cognition}, 7\penalty0 (4):\penalty0 409--411, 1979.

\bibitem[Latham and Locke(2007)]{latham2007new}
G.~P. Latham and E.~A. Locke.
\newblock New developments in and directions for goal-setting research.
\newblock \emph{European Psychologist}, 12\penalty0 (4):\penalty0 290, 2007.

\bibitem[Levy et~al.(2017)Levy, Sarne, and Rochlin]{levy2017contest}
P.~Levy, D.~Sarne, and I.~Rochlin.
\newblock Contest design with uncertain performance and costly participation.
\newblock In \emph{Proceedings of the 26th International Joint Conference on
  Artificial Intelligence (IJCAI)}, pages 302--309, 2017.

\bibitem[Main and Ost(2014)]{Main14}
J.~B. Main and B.~Ost.
\newblock The impact of letter grades on student effort, course selection, and
  major choice: A regression-discontinuity analysis.
\newblock \emph{The Journal of Economic Education}, 45\penalty0 (1):\penalty0
  1--10, 2014.

\bibitem[McEwan et~al.(2021)McEwan, Rogers, and Weerapana]{mcewan2021grade}
P.~J. McEwan, S.~Rogers, and A.~Weerapana.
\newblock Grade sensitivity and the economics major at a women's college.
\newblock In \emph{AEA papers and proceedings}, volume 111, pages 102--06,
  2021.

\bibitem[Nisbet(1975)]{nisbet1975adding}
J.~Nisbet.
\newblock Adding and averaging grades.
\newblock \emph{Educational Research}, 17\penalty0 (2):\penalty0 95--100, 1975.

\bibitem[Paredes(2017)]{parades17}
V.~Paredes.
\newblock Grading system and student effort.
\newblock \emph{Education Finance and Policy}, 12\penalty0 (1):\penalty0
  107--128, 2017.

\bibitem[Rohe et~al.(2006)Rohe, Barrier, Clark, Cook, Vickers, and
  Decker]{db2006}
D.~Rohe, P.~Barrier, M.~Clark, D.~Cook, K.~Vickers, and P.~A. Decker.
\newblock The benefits of pass-fail grading on stress, mood, and group cohesion
  in medical students.
\newblock \emph{Mayo Clinic Proceedings}, 81\penalty0 (11):\penalty0
  1443--1448, 2006.

\bibitem[Sikora(2015)]{Sikora2015}
A.~S. Sikora.
\newblock Mathematical theory of student assessment through grading.
\newblock
  \url{http://citeseerx.ist.psu.edu/viewdoc/download?doi=10.1.1.714.8666&rep=rep1&type=pdf},
  2015.
\newblock Accessed: 2022-08-15.

\bibitem[{University of Waterloo}(2022)]{nonmidpoint}
{University of Waterloo}.
\newblock {University of Waterloo} graduate studies, grading scheme prior to
  {F}all 2001.
\newblock
  \url{https://uwaterloo.ca/graduate-studies-academic-calendar/general-information-and-regulations/grades-and-grading},
  2022.
\newblock Accessed: 2022-08-15.

\bibitem[{University of Western Ontario}(2022)]{midpoint1}
{University of Western Ontario}.
\newblock {University of Western Ontario} grading scheme.
\newblock
  \url{https://registrar.uwo.ca/academics/grades_progression_and_graduation.html#gpa},
  2022.
\newblock Accessed: 2022-08-15.

\bibitem[{Victoria University of Wellington}(2022)]{midpoint2}
{Victoria University of Wellington}.
\newblock {Victoria University of Wellington} grading scheme.
\newblock \url{https://www.wgtn.ac.nz/students/study/progress/grades}, 2022.
\newblock Accessed: 2022-08-15.

\bibitem[Wedell et~al.(1989)Wedell, Parducci, and Roman]{wedell1989student}
D.~H. Wedell, A.~Parducci, and D.~Roman.
\newblock Student perceptions of fair grading: A range-frequency analysis.
\newblock \emph{The American Journal of Psychology}, pages 233--248, 1989.

\bibitem[Zavaleta-Bernuy et~al.(2021)Zavaleta-Bernuy, Zheng, Shaikh, Nogas,
  Rafferty, Petersen, and Williams]{zavaleta2021using}
A.~Zavaleta-Bernuy, Q.~Y. Zheng, H.~Shaikh, J.~Nogas, A.~Rafferty, A.~Petersen,
  and J.~J. Williams.
\newblock Using adaptive experiments to rapidly help students.
\newblock In \emph{Proceedings of the 22nd International Conference on
  Artificial Intelligence in Education (AIED)}, pages 422--426, 2021.

\end{thebibliography}

\appendix
\onecolumn

\section{Intuition Regarding \texorpdfstring{$\calD^{\same}$}{Dsame} vs \texorpdfstring{$\calD^{\opp}$}{Dopp} \& Single-Peakedness}\label{app:intuition}

\pgfmathdeclarefunction{gauss}{2}{%
  \pgfmathparse{1/(#2*sqrt(2*pi))*exp(-((x-#1)^2)/(2*#2^2))}%
}
\pgfplotsset{
  /pgfplots/xlabel near ticks/.style={
     /pgfplots/every axis x label/.style={
        at={(ticklabel cs:0.5)},anchor=near ticklabel
     }
  },
  /pgfplots/ylabel near ticks/.style={
     /pgfplots/every axis y label/.style={
        at={(ticklabel cs:0.5)},rotate=90,anchor=near ticklabel
     }
  },
  major grid style={dashed,gray!70!black},
  xtick align=outside
}
  
\begin{figure}[htb!]
\centering
    \begin{subfigure}[t]{0.8\textwidth}
        \begin{tikzpicture}
            \begin{axis}[
                  no markers, domain=68:87, samples=100,
                  axis lines*=left, xlabel=Score $s$, ylabel=Density $f_{\calS}(s;73)$,
                  height=5cm, width=12cm,
                  xtick={60,65,73,70,75,80,85}, ytick=\empty,
                  enlargelimits=false, clip=false, axis on top,
                  grid = minor
                  ]
                 \node[coordinate, pin={$\calD^{\same}$}] at (axis cs: 71.5, 0.16){};
                  
                 \node[coordinate, pin={$\calD^{\opp}$}] at (axis cs: 76, 0.05){};
                  
            	\addplot [fill=cyan!70!black,  domain=70:75] {gauss(73,1.7)} \closedcycle;
                \addplot [fill=red!70!black,  domain=75:80] {gauss(73,1.7)} \closedcycle;
                 \draw [dashed,gray!70!black] (axis cs: 73,0) -- (axis cs: 73,0.236);
            \end{axis}
        \end{tikzpicture}
        \caption{With a small value of $\gamma$, one can see that within the grade interval $[70,80]$ containing the true quality $q=73$, the probability of the score being on the same side of the midpoint as the true quality (i.e., in $[70,75]$) is significantly higher than the probability of it being on the opposite side of the midpoint (i.e., in $[75,80]$). The former region contributes to $\calD^{\same}$ while the latter contributes to $\calD^{\opp}$. Their difference is the most pronounced when the true quality is near the interval endpoints (e.g., $q \approx 70,80$) and gradually vanishes when it is near the midpoint (e.g., $q \approx 75$). In expectation over  the true quality, one can still expect $\Pr[(q,s) \in \calD^{\same}]$ to be sufficiently higher than $\Pr[(q,s) \in \calD^{\opp}]$, satisfying the conditions in \Cref{thm:weak-symm-impr-2} and \Cref{clm:threetimesbucketing}.}
    \end{subfigure}\\[0.5cm]%
    \begin{subfigure}[t]{0.8\textwidth}
        \begin{tikzpicture}
            \begin{axis}[
              no markers, domain=58:94, samples=100,
              axis lines*=left,  xlabel=Score $s$, ylabel=Density $f_{\calS}(s;73)$,
              height=5cm, width=12cm,
              xtick={60,65,70, 73,75, 80,85,90}, ytick=\empty,
              enlargelimits=false, clip=false, axis on top,
              grid = minor
              ]
            
        	\addplot [fill=gray!30!white,  domain=60:65] {gauss(73,6)} \closedcycle;
        	\addplot [fill=gray!70!white,  domain=65:70] {gauss(73,6)} \closedcycle;
  	
        	\addplot [fill=gray!70!white,  domain=80:85] {gauss(73,6)} \closedcycle;
        	\addplot [fill=gray!30!white,  domain=85:90] {gauss(73,6)} \closedcycle;
 
        	\addplot [very thick,gray!70!black,  domain=60:90] {gauss(73,6)} \closedcycle;
            \draw [dashed,gray!70!black] (axis cs: 73,0) -- (axis cs: 73,0.067);
            \end{axis}
        \end{tikzpicture}
        \caption{Due to single-peakedness of the score distribution, the expected score in any interval lower than the interval containing the true quality $q=73$ is at least its midpoint (e.g., the expected score subject to the score being in $[60,70]$ is at least $65$). In contrast, the expected score in any interval higher than the interval containing the true quality $q=73$ is at most its midpoint (e.g., the expected score subject to the score being in $[80,90]$ is at most $85$). This observation is used at the end of the proof of \Cref{thm:weak-symm-impr}.}
    \end{subfigure}
    \caption{Both figures show the probability density function of the score distribution $\calS(q)$ when the true quality is $q=73$. The distribution is a truncated normal distribution with mean $q=73$, and standard deviation $\gamma = 1.7$ (top figure) and $\gamma = 6$ (bottom figure). The top figure conveys the intuition behind the conditions in \Cref{thm:weak-symm-impr-2} and \Cref{clm:threetimesbucketing}, which assume $\Pr[(q,s) \in \calD^{\same}]$ to be sufficiently higher than $\Pr[(q,s) \in \calD^{\opp}]$. The bottom figure conveys the intuition behind the observation used at the end of the proof of \Cref{thm:weak-symm-impr}.}
    \label{fig:intuition}
\end{figure}

\section{Useful Lemmas}\label{app:lemmas}
Before we dive into the missing proofs, we state the integral version of the well-known Chebyshev's inequality and its two useful implications. 

\begin{lemma}[Integral Chebyshev Inequality]\label{lem:cheb}
If functions $f,g:[a,b] \to \bbR_{\ge 0}$ are either both non-increasing or both non-decreasing, then 
\begin{align*}
    \frac{1}{b-a} \int_a^b f(x) g(x) \dif x \ge \left( \frac{1}{b-a} \int_a^b f(x) \dif x \right) \cdot \left( \frac{1}{b-a} \int_a^b g(x) \dif x \right).
\end{align*}
If one of them is non-decreasing while the other is non-increasing, the inequality is reversed.
\end{lemma}

The following inequality is obtained by substituting $f(x) = x$ (and thus, $\frac{1}{b-a} \int_a^b f(x) \dif x = \frac{a+b}{2}$) into \Cref{lem:cheb} from \Cref{app:lemmas}.

\begin{lemma}\label{lem:cheb2}
If $g:[a,b] \to \bbR_{\ge 0}$ is a non-increasing function, then we have
\[
\int_a^b x g(x) \dif x \le \frac{a+b}{2} \cdot \int_a^b g(x) \dif x,
\]
and the inequality is reversed if $g$ is a non-decreasing function.
\end{lemma}

If $g$ is a probability density function over $[a,b]$, then $\int_a^b g(x) \dif x = 1$, yielding the following (quite natural) implication.

\begin{lemma}\label{lem:cheb-exp}
Let $X$ be a random variable over $[a,b]$ with a non-increasing probability density function $g:[a,b] \to \bbR_{\ge 0}$. Then, $\E[X] \le (a+b)/2$, and the inequality is reversed if $g$ is non-decreasing.
\end{lemma}

Finally, we use the following strengthening of the integral Chebyshev inequality when one of the functions is linear and the other is concave non-increasing.

\begin{lemma}\label{lem_fd_last}
Let $g : [a,b] \to \bbR_{\ge 0}$ be a concave function with $g(b)=0$. Then, we have 
\[
\int_a^b (b-x) g(x) \dif x \leq \frac{2(b-a)}{3} \int_a^b g(x) \dif x.
\]
\end{lemma}
\begin{proof}
Due to concavity of $g$, we have
\[
\int_x^b g(t) \dif t  \ge \frac{1}{2} (b-x) g(x). 
\]

Hence, we have
\begin{align*}
\int_a^b \frac{1}{2} (b-x) g(x) \dif x &\ge \int_{x=a}^b \int_{t=x}^b g(t) \dif t \dif x\\
&= \int_{t=a}^b \int_{x=a}^t g(t) \dif x \dif t &&\text{(Fubini's theorem)}\\
&= \int_{t=a}^b (t-a) g(t) \dif t\\
&=\int_{x=a}^b (x-a) g(x) \dif x &&\text{(Change of variable name)}\\
&=\int_a^b (b-a) g(x) \dif x - \int_a^b (b-x) g(x) \dif x.
\end{align*}
Rearranging the terms yields the desired inequality. 
\end{proof}

\section{Additional Experimental Results}\label{app:more-figures}

In our experiments, we compared numerical scoring to uniform letter grading schemes with $T \in \set{4,8,12,16,20}$ grades. In the main text, we presented results that show the impact of two parameters, the number of evaluations $r$ and the motivation coefficient $\alpha_m$, when one of them is varied while keeping the other fixed. 

\begin{figure*}[ht!]
    \centering
    \begin{subfigure}[t]{0.47\textwidth}
        \includegraphics[width=\textwidth]{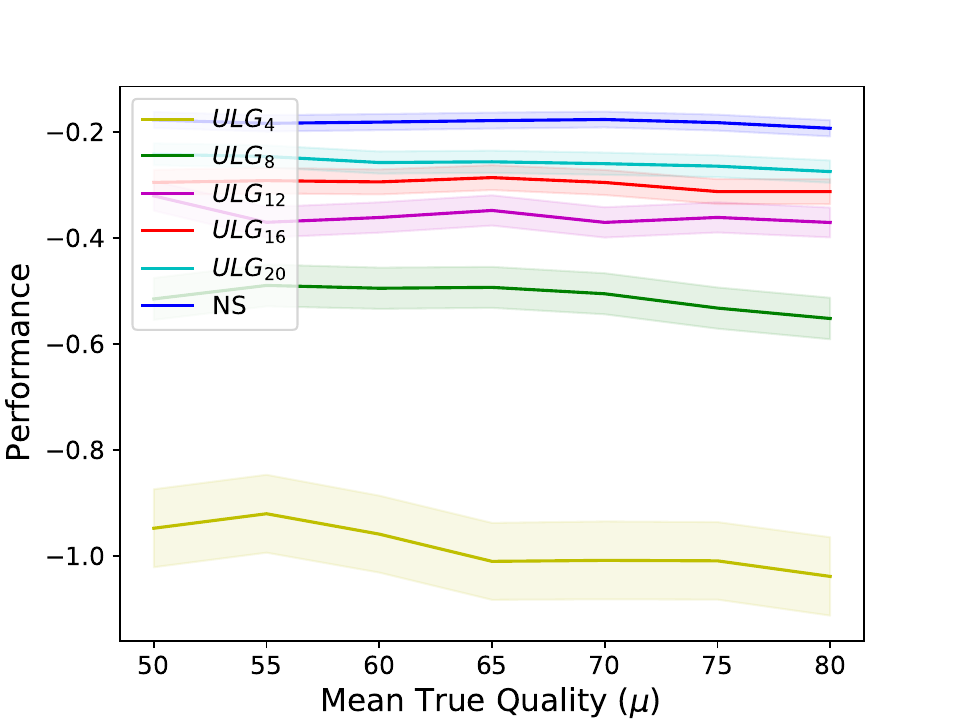}
        \caption{\textsc{$\alpha_m=0.2, r=2$}}
        \label{subfig:mu-final-0.2}
    \end{subfigure}\hspace{0.05\textwidth}%
    \begin{subfigure}[t]{0.47\textwidth}
        \centering
        \includegraphics[width=\textwidth]{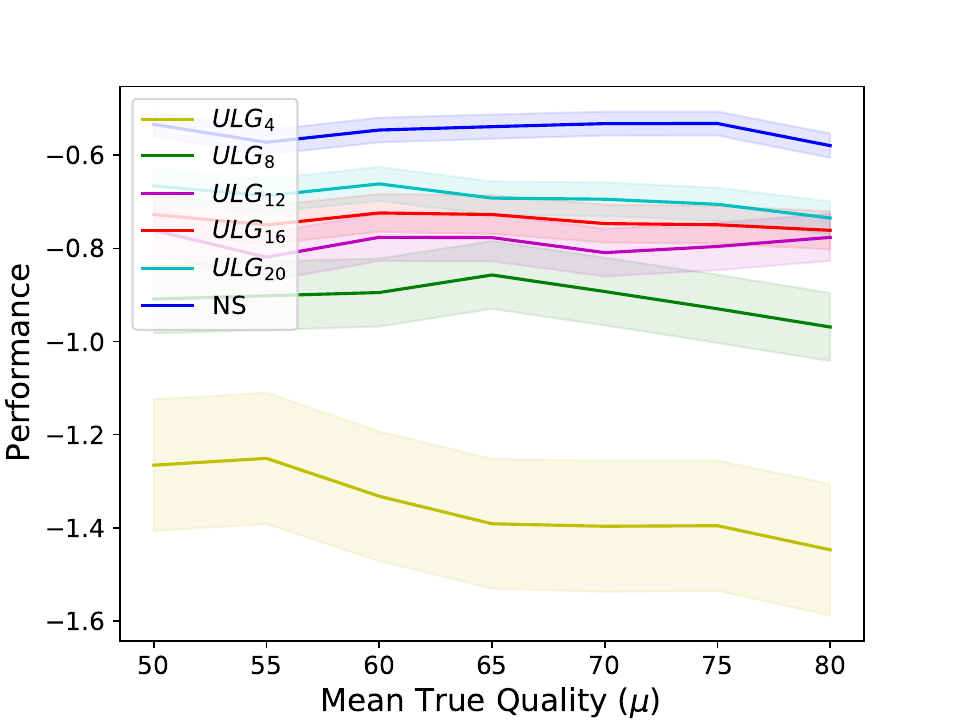}
        \caption{$\alpha_m=0.2, r=4$}
        \label{subfig:mu-final-0.8}
    \end{subfigure}
    \begin{subfigure}[t]{0.47\textwidth}
        \includegraphics[width=\textwidth]{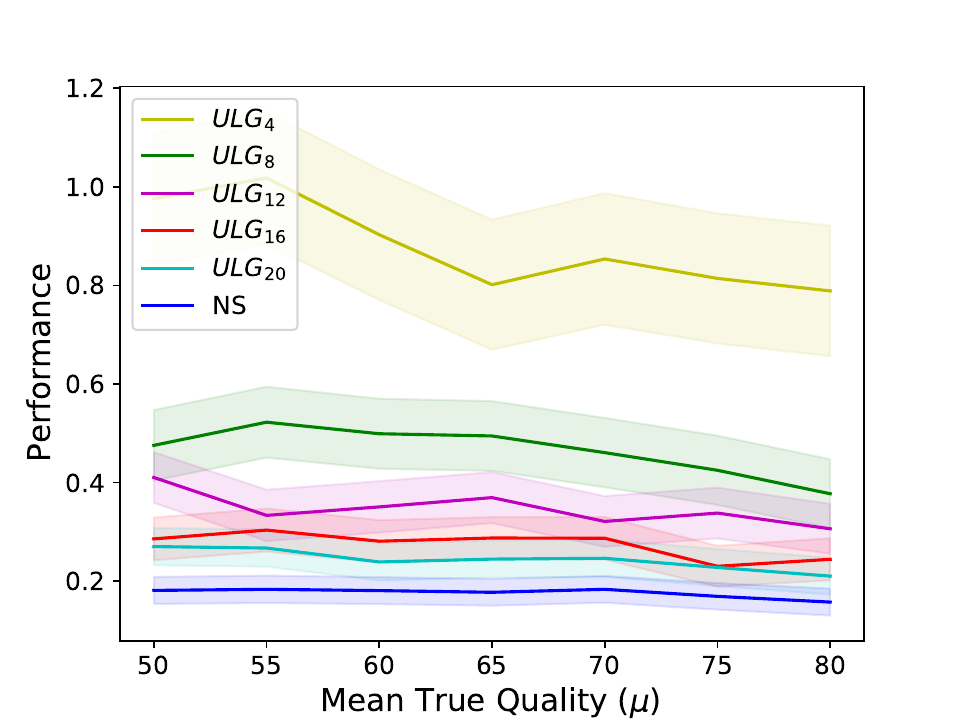}
        \caption{\textsc{$\alpha_m=0.8, r=2$}}
        \label{subfig:mu-avg-0.2}
    \end{subfigure}\hspace{0.05\textwidth}%
    \begin{subfigure}[t]{0.47\textwidth}
        \centering
        \includegraphics[width=\textwidth]{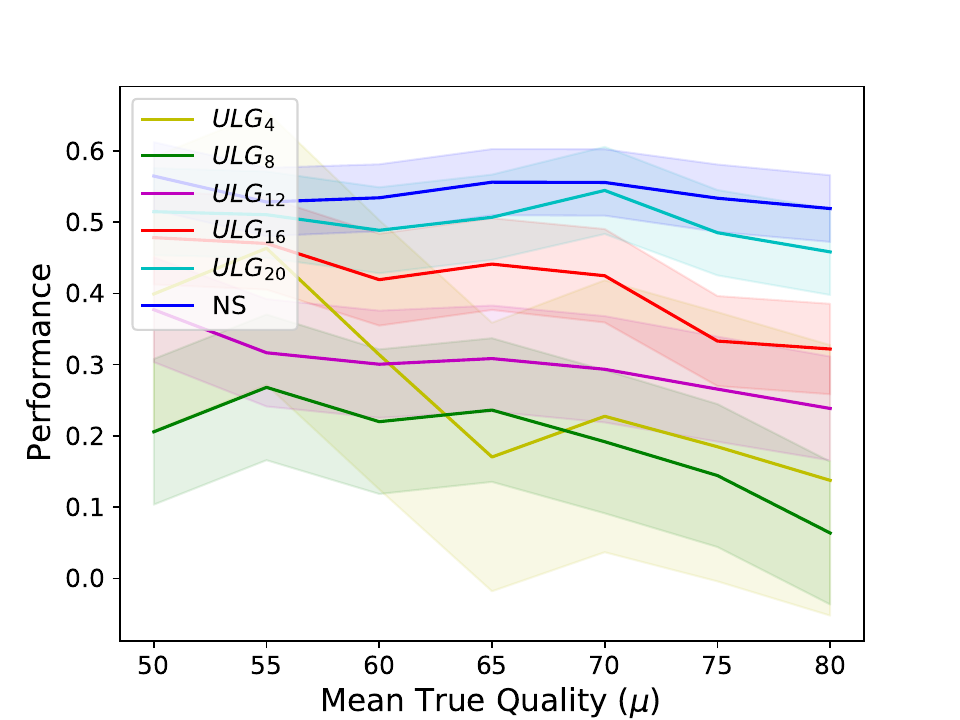}
        \caption{$\alpha_m=0.8, r=4$}
        \label{subfig:mu-avg-0.8}
    \end{subfigure}
     \caption{ Performance of numerical scoring and different uniform letter grading schemes, with  $\sigma=12$,  $\gamma=1.5$ and $\alpha_d=0.5$, over different values of $\mu$.  }
    \label{fig:mu}
\end{figure*}

\begin{figure*}[ht!]
    \centering
    \begin{subfigure}[t]{0.47\textwidth}
        \includegraphics[width=\textwidth]{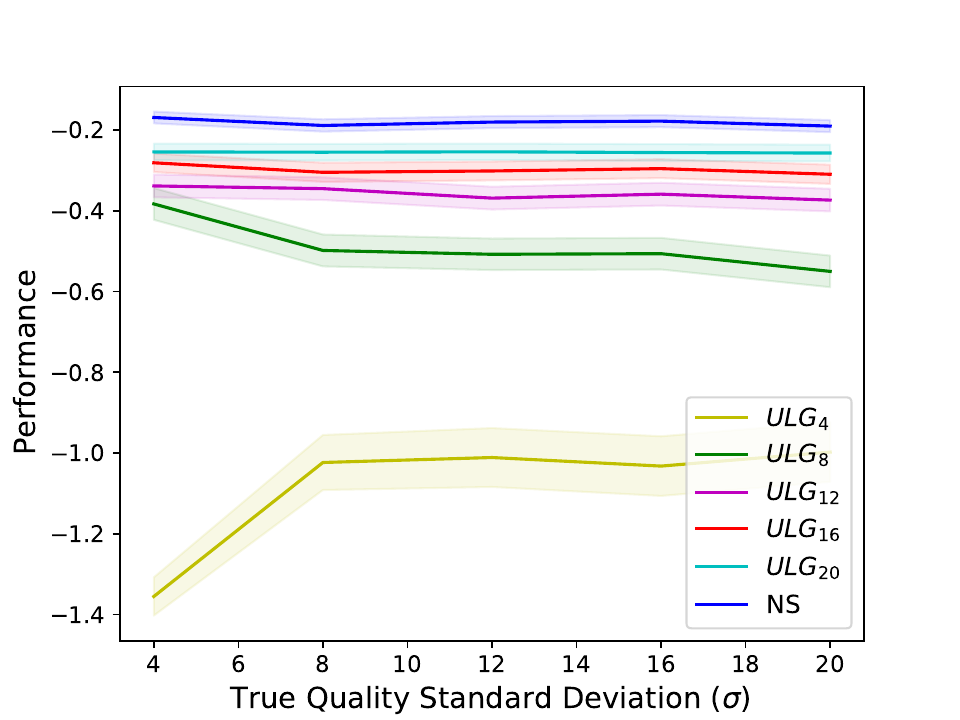}
        \caption{\textsc{$\alpha_m=0.2, r=2$}}
        \label{subfig:sigma-final-0.2}
    \end{subfigure}\hspace{0.05\textwidth}%
    \begin{subfigure}[t]{0.47\textwidth}
        \centering
        \includegraphics[width=\textwidth]{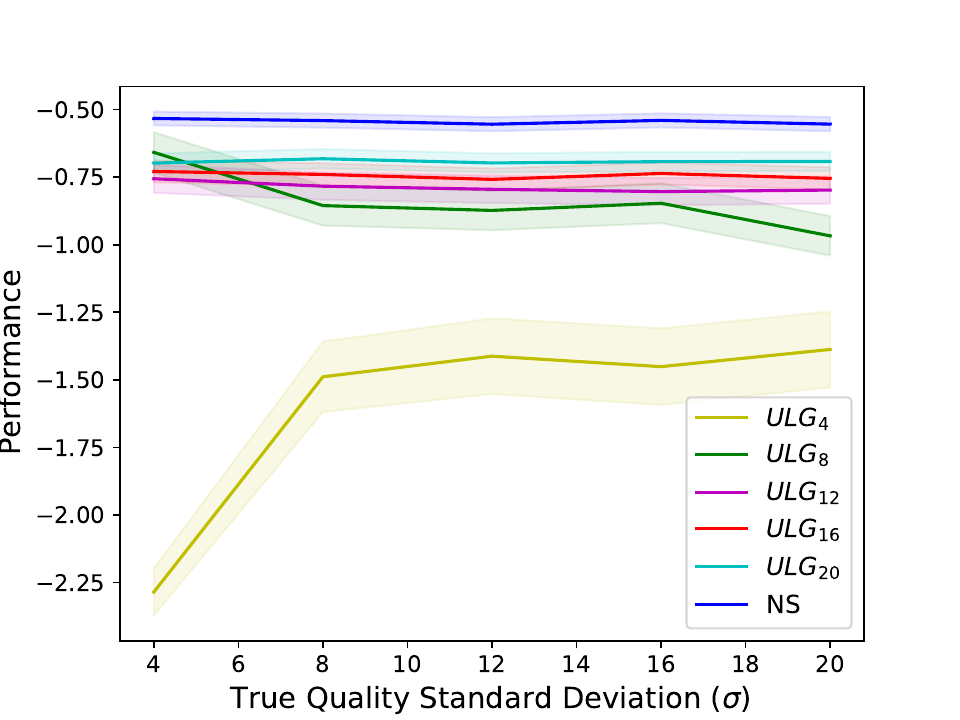}
        \caption{$\alpha_m=0.2, r=4$}
        \label{subfig:sigma-final-0.8}
    \end{subfigure}
    \begin{subfigure}[t]{0.47\textwidth}
        \includegraphics[width=\textwidth]{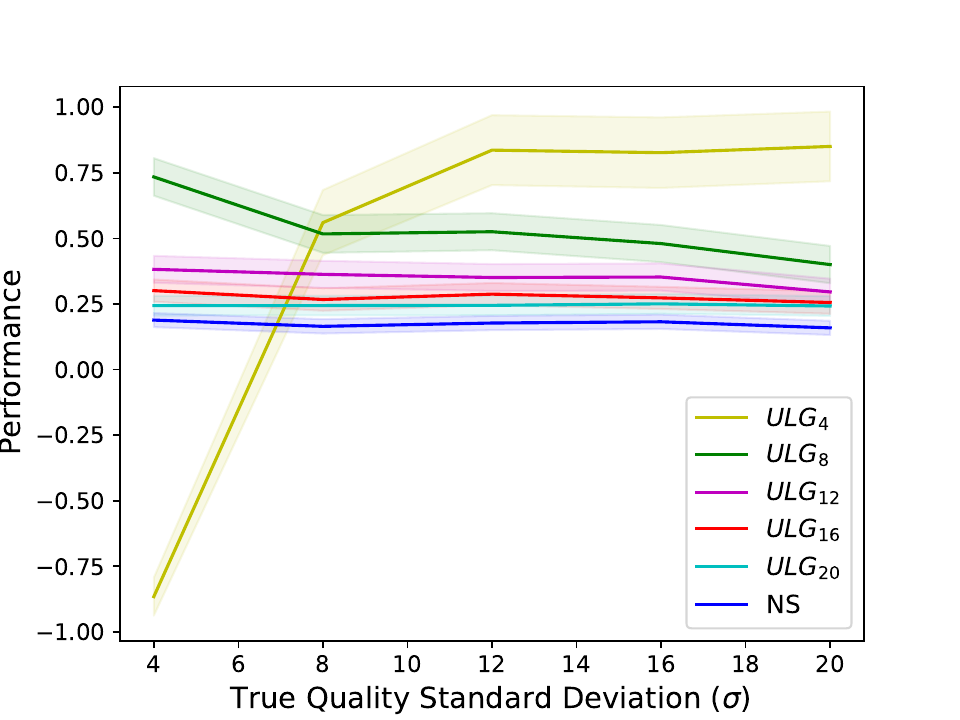} 
        \caption{\textsc{$\alpha_m=0.8, r=2$}}
        \label{subfig:sigma-avg-0.2}
    \end{subfigure}\hspace{0.05\textwidth}%
    \begin{subfigure}[t]{0.47\textwidth}
        \centering
        \includegraphics[width=\textwidth]{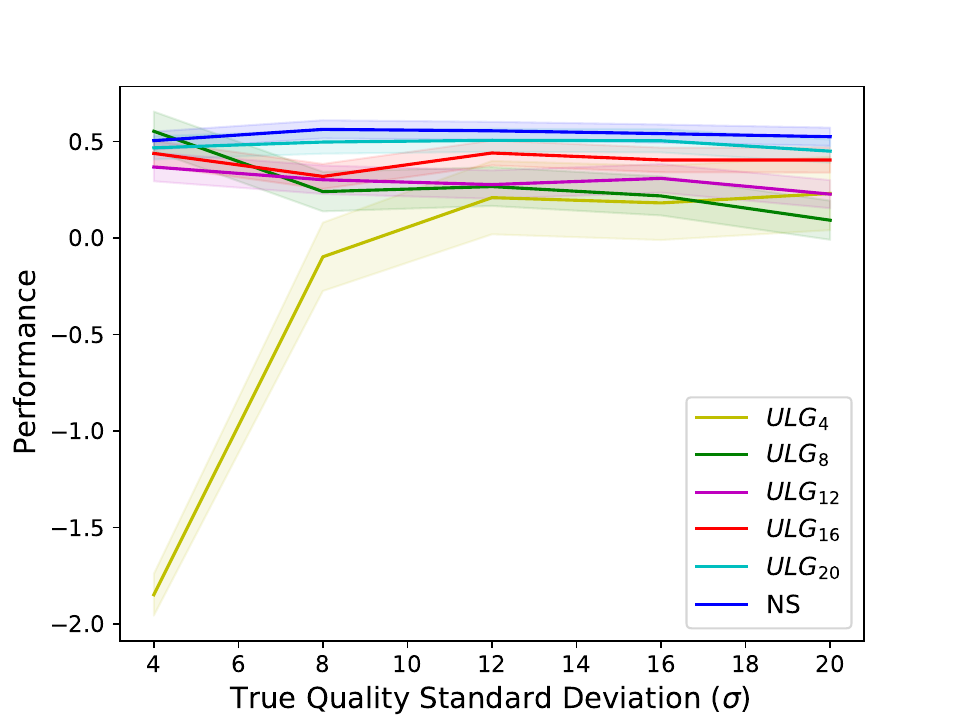}
        \caption{$\alpha_m=0.8, r=4$}
        \label{subfig:sigma-avg-0.8}
    \end{subfigure}
     \caption{ Performance of numerical scoring and different uniform letter grading schemes, with  $\mu=65$,  $\gamma=1.5$ and $\alpha_d=0.5$, over different values of $\sigma$.  }
    \label{fig:sigma}
\end{figure*}

\begin{figure*}[ht!]
    \centering
    \begin{subfigure}[t]{0.47\textwidth}
        \includegraphics[width=\textwidth]{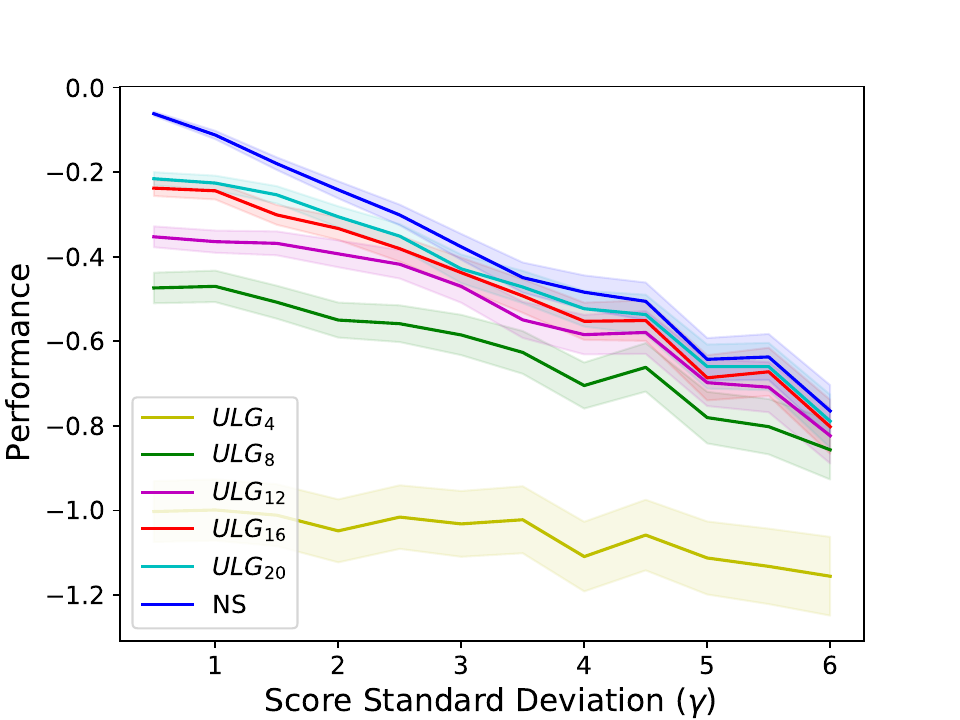}
        \caption{\textsc{$\alpha_m=0.2, r=2$}}
        \label{subfig:gamma-final-0.2}
    \end{subfigure}\hspace{0.05\textwidth}%
    \begin{subfigure}[t]{0.47\textwidth}
        \centering
        \includegraphics[width=\textwidth]{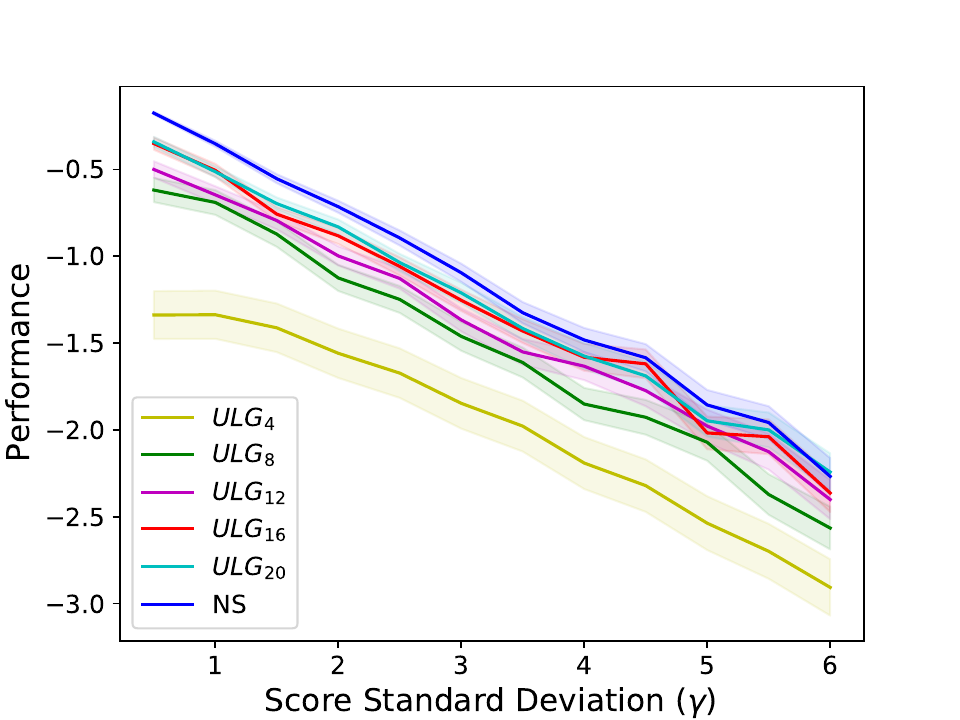}
        \caption{$\alpha_m=0.2, r=4$}
        \label{subfig:gamma-final-0.8}
    \end{subfigure}
    \begin{subfigure}[t]{0.47\textwidth}
        \includegraphics[width=\textwidth]{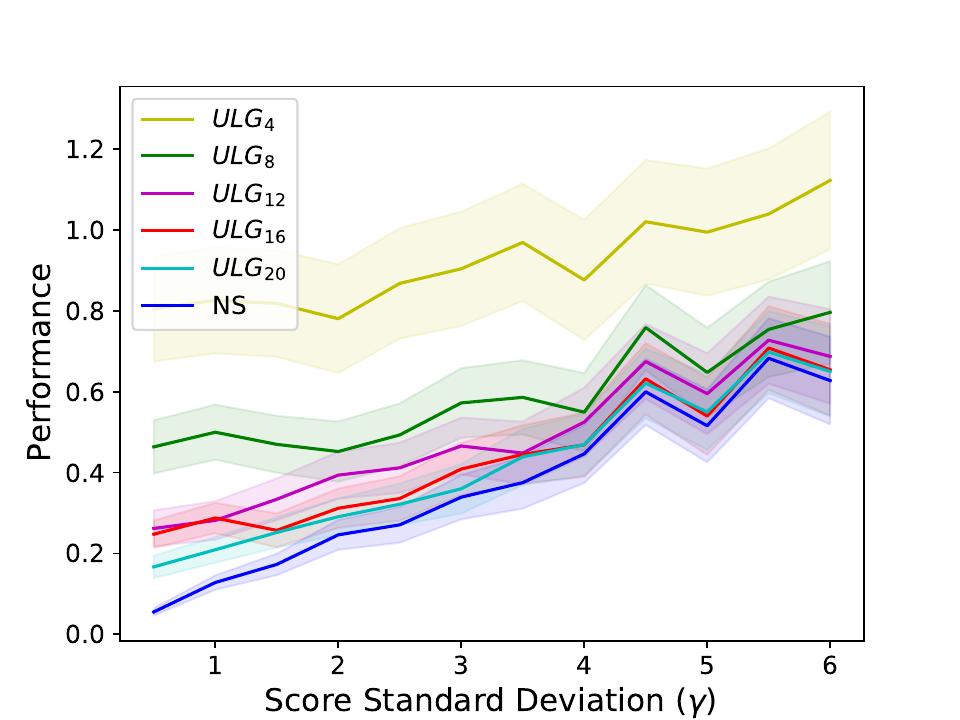} 
        \caption{\textsc{$\alpha_m=0.8, r=2$}}
        \label{subfig:gamma-avg-0.2}
    \end{subfigure}\hspace{0.05\textwidth}%
    \begin{subfigure}[t]{0.47\textwidth}
        \centering
        \includegraphics[width=\textwidth]{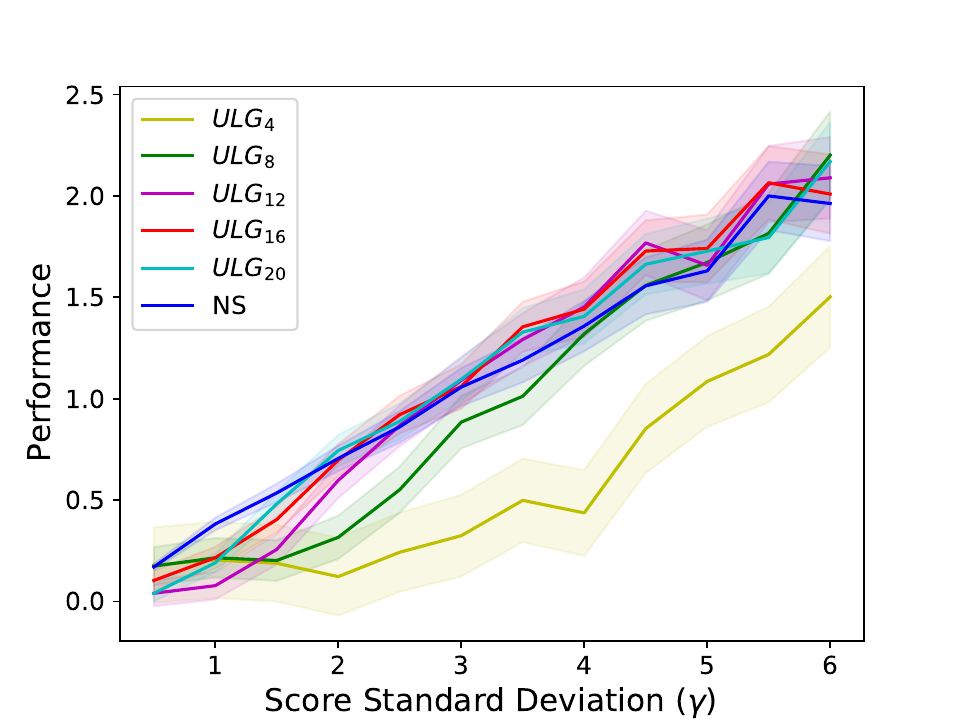}
        \caption{$\alpha_m=0.8, r=4$}
        \label{subfig:gamma-avg-0.8}
    \end{subfigure}
     \caption{ Performance of numerical scoring and different uniform letter grading schemes, with  $\mu=65$,  $\sigma=12$ and $\alpha_d=0.5$, over different values of $\gamma$.  }
    \label{fig:gamma}
\end{figure*}

Here, we present additional experimental results, which show the impact of varying the mean $\mu$ of the true quality distribution (\Cref{fig:mu}), the standard deviation $\sigma$ of the true quality distribution (\Cref{fig:sigma}), and the standard deviation $\gamma$ of the score distribution (\Cref{fig:gamma}).\footnote{Technically, these are the mean and the standard deviations of the respective underlying normal distributions before truncation.} 

Overall, the mean true quality $\mu$ has little impact on the performance of   different grading schemes. Similarly, the standard deviation $\sigma$ of the true quality prior also does not significantly affect the performance of the grading schemes, but somewhat strikingly, it has a dramatic impact on the performance of $\ulg_4$ (uniform letter grading with $4$ grades).

The impact of  the standard deviation $\gamma$ of the score distribution is more significant, since as $\gamma$ increases, the performance of the different grading schemes becomes more similar. However, we see that even for quite large values of  of $\gamma$,   our theoretical results seem to hold. In particularly, we see that when two evaluations are taken place, numerical scoring is better when $\alpha_m < \alpha_d$ whereas uniform letter grading is better when $\alpha_m > \alpha_d$. This observation is quite encouraging since it probably indicates that our results can be extended for cases where the score is not very well-concentrated around the true quality.

\end{document}